\documentclass[11pt,a4paper]{article}
\usepackage{jcappub}

\usepackage{appendix}
\usepackage{array}
\newcolumntype{x}[1]{>{\centering\arraybackslash}p{#1}}

\def\lsim{\mathrel{\hbox{\rlap{\hbox{\lower4pt\hbox{$\sim$}}}\hbox{$<$}}}}

\makeatletter

\newcommand{\Rmnum}[1]{\expandafter\@slowromancap\romannumeral #1@}
\makeatother

\title{Ring-like features in directional dark matter detection}

\author[a]{Nassim Bozorgnia,}
\author[a]{Graciela B. Gelmini}
\author[b,c]{and Paolo Gondolo}
\affiliation[a]{Department of Physics and Astronomy, UCLA,\\
475 Portola Plaza, Los Angeles, CA 90095, USA}
\affiliation[b]{Department of Physics and Astronomy, University of Utah,\\
115 South 1400 East \#201, Salt Lake City, UT 84112, USA}
\affiliation[c]{Stockholm University, The Oskar Klein Centre,\\
SE-106 91 Stockholm, Sweden}
\emailAdd{nassim@physics.ucla.edu}
\emailAdd{gelmini@physics.ucla.edu}
\emailAdd{paolo@physics.utah.edu}

\abstract{
We discuss  a novel dark matter signature  relevant for directional detection of Weakly Interacting Massive Particles (WIMPs). For heavy enough WIMPs and low enough recoil energies, the maximum of the recoil rate is not in the direction of  the average WIMP arrival direction but in a ring around it at an angular radius that increases with the WIMP mass and can approach 90$^\circ$ at very low energies. The ring is easier to detect for smaller WIMP velocity dispersion and larger  average WIMP velocities relative to the detector. In principle the ring could be used as an additional indication of the WIMP mass range.}

\arxivnumber{1111.6361}

\begin{document}
\maketitle

\section{Introduction}

Direct dark matter experiments search for energy deposited in low-background detectors by the scattering of Weakly Interacting Massive Particles (WIMPs) present in the dark halo of our galaxy. If a signal is observed, it is necessary to have unmistakable signatures that it is due to dark matter. Two such signatures are the annual modulation of the signal due to the motion of the Earth around the Sun, extensively studied since~\cite{Drukier}, and the daily modulation due to the spinning of the Earth around its axis.

With the advent of directional detectors~\cite{Ahlen:2009ev} using CS$_2$, CF$_4$ or $^3$He, such as DRIFT~\cite{DRIFT}, DMTPC~\cite{DMTPC}, NEWAGE~\cite{NEWAGE} and MIMAC~\cite{MIMAC}, the observation of other dark matter signatures becomes possible.  Because dark matter WIMPs arrive to us from a preferential direction, one expects an anisotropy in the recoil event rate~\cite{Spergel}. The detector is moving with a velocity ${\bf V}_{\rm {lab}}$ with respect to the Galaxy. If the particles in the dark halo of our galaxy are on average at rest with respect to the Galaxy, the average velocity of the dark matter particles with respect to the detector is $-{\bf V}_{\rm {lab}}$. This is the case in the standard halo model. Here we consider a galactic dark matter halo model whose velocity distribution, as in the standard halo model, is locally approximated  by  an isotropic  Maxwell-Boltzmann (IMB) distribution, but with a velocity dispersion  independent of the average velocity, to better represent what is known about the actual local parameters of the distribution. We use the IMB  in most of this paper.

The easiest dark matter signature to detect in directional detectors is the average recoil direction, expected to be in the direction of $-{\bf V}_{\rm lab}$~\cite{Spergel, Copi:1999pw,Morgan:2005,Green:2010gw}.  Ref.~\cite{Green:2010gw} finds  that  for an S detector with an energy threshold of 20 keV, 9 events would be enough to reject isotropy at 95\% CL, while between 27 and 32 events, depending on the velocity distribution, would be enough  to confirm the direction of solar motion as the median inverse recoil direction at 95\% CL. For a 100 GeV/$c^2$ WIMP, with spin-independent interactions and WIMP-proton cross section $\sigma_p=10^{-44}\;{\text{cm}}^2$, 30 events in CS$_2$ corresponds to an exposure of 150 kg-yr~\cite{Morgan:2005}. Ref.~\cite{Green:2010gw} assumes the recoil directions, including their senses, can be reconstructed perfectly in 3d and the background is zero. Assuming a less optimistic model for a CF$_4$ detector with a 50\% background contamination, Ref.~\cite{Billard:2009mf} finds that a signal pointing within 20$^{\circ}$ of the opposite direction to the Solar motion can be confirmed with 25 events between 5 keV and 50 keV.

In this paper we discuss an additional dark matter signature relevant for directional detection as a secondary feature with respect to the mean recoil direction: a ring of  maximum recoil rate around the direction of $-{\bf V}_{\rm lab}$. For heavy enough WIMPs and low enough energies, the maximum of the recoil rate is not in the direction of  the average WIMP arrival direction but in a ring around it at an angle which can be up to 90$^{\circ}$.

  \begin{figure}[t]
\begin{center}
  \includegraphics[height=170pt]{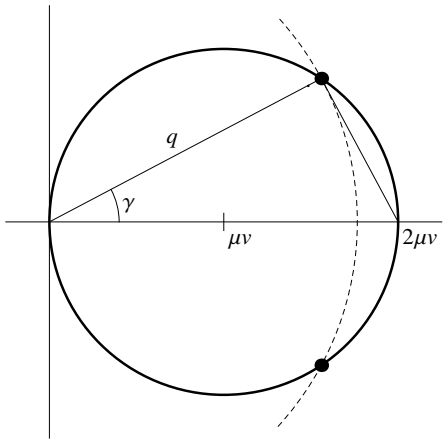}\\
  \vspace{-0.1cm}\caption{Figure describing the existence of a ring in recoil ${\bf q}$ space due to a WIMP, or  WIMP stream, with velocity ${\bf v}$. $\mu$ is the reduced WIMP-nucleus mass.}
  \label{Ring-Explain}
\end{center}
\end{figure}

The origin of the ring can be understood considering just one incoming WIMP with velocity $\bf{v}$. Energy and momentum conservation in the collision of this WIMP with a nucleus of mass $M$ imply that the magnitude of the recoil nuclear momentum is $q=2\mu v \cos \theta$, where $v=|{\bf v}|$, $\mu$ is the reduced WIMP-nucleus mass and $\theta$ is the scattering angle, i.e.\ the angle between ${\bf v}$ and the recoil momentum ${\bf q}$~\cite{Gondolo:2002}. As shown in Fig.~\ref{Ring-Explain}, in momentum space these values of $q$ lie on the surface of a sphere of radius $\mu v$, centered at $\mu {\bf v}$ and passing through the origin. For a certain recoil energy $E_R=q^2/2M$, ${\bf q}$ must be, in momentum space, on a sphere with radius $q=\sqrt{2M E_R}$ and centered at the origin ${\bf q}=0$. The only possible ${\bf q}$ values at fixed magnitude $q$ are therefore on the ring on which the two spheres mentioned intersect. The angular radius of the ring $\gamma$ fulfills the condition $\cos \gamma=\sqrt{2M E_R}/2\mu v$. In the IMB, most WIMPs arrive with velocity $-{\bf V}_{\rm {lab}}$, so the maximum number of recoils occur when $v=V_{\rm lab}$ (see Eq.~\ref{gamma}). As $V_{\rm {lab}}$ increases, the angle $\gamma$ increases, and the flux in the direction of $-{\bf V}_{\rm {lab}}$ decreases with respect to the maximum rate. This makes the ring easier to observe.

Our ring-like signature would require more events than those mentioned above to be detected. In Section 3 we find with a simple statistical test that for a CS$_2$ detector with 3d track reconstruction and a 100 GeV/$c^2$ WIMP, for recoil energies between 5 keV and 40 keV and no background, 41 events total over the whole sky would be required to observe the ring in one of the most favorable cases (not taking into account the energy and angular experimental resolutions). This corresponds to an exposure of 250 kg-yr for $\sigma_p=10^{-44}\;{\text{cm}}^2$. For recoil energies between 20 keV and 40 keV instead, 55 events in total over the whole sky would be required, corresponding to 700 kg-yr for $\sigma_p=10^{-44}\;{\text{cm}}^2$. These are lower bounds on the number of events needed to detect the ring.

We noticed the existence of this ring before (see e.g.\ Fig.~4 of Ref.~\cite{ChanIV} and the explanation of this figure) but to our knowledge its potential relevance for directional detection has not been pointed out.

\section{Differential Recoil Spectrum}

It is easy to see that at low recoil energies for heavy enough WIMPs the maximum of the directional recoil rate is not in the direction of $-{\bf V}_{\rm lab}$, by writing the recoil spectrum in terms of the Radon transform. Let us start with the directional differential recoil spectrum as a function of the recoil momentum ${\bf q}$~\cite{Alenazi-Gondolo:2008}
\begin{equation}
\frac{dR}{ dE_R~d\Omega_q}= \sum_i \frac{\rho}{4\pi m \mu_i^2} C_i \hat{f}_{\rm lab}\!\left( \frac{q}{2\mu_i}, \hat{\bf q} \right) \sigma_i(q).
\label{eq: rate}
\end{equation}
Here $\rho$ is the dark matter density in the solar neighborhood and $m$ is the WIMP mass. The sum is over the nuclear species $i$  in the target, and $C_i$ and $\mu_i$ are the mass fraction and the reduced WIMP-nucleus mass for nuclide $i$, respectively. $d\Omega_q=d\phi \, d\!\cos\theta$ denotes an infinitesimal solid angle around the recoil direction $\hat{\bf q}= {\bf q}/q$, $q=|{\bf q}|$ is the magnitude of the recoil momentum, $q/2\mu=v_q$ is the minimum velocity a WIMP must have to impart a recoil momentum $q$ to the nucleus of mass $M$, or equivalently to deposit a recoil energy $E_R = q^2/2M$,   $\mu=m M/(m+M)$, and $\hat{ f}_{\rm lab}$ is the 3-dimensional Radon transform of the WIMP velocity distribution. $\sigma_i(q)$ is the WIMP-nucleus scattering cross section which can be split into spin-independent (SI) and spin-dependent (SD) parts, $\sigma_i(q)=\sigma_i^{\rm SI}(q) +\sigma_i^{\rm SD}(q)$.

In the usual IMB model, WIMPs are on average at rest with respect to the Galaxy, and have a Maxwellian velocity distribution with dispersion $\sigma_v$, truncated at the escape speed $v_{\rm esc}$ (with respect to the Galaxy). Normalized to 1, the WIMP velocity distribution in the laboratory rest-frame is given by~\cite{Gondolo:2002}
\begin{equation}
f_{\rm WIMP}({\bf v})=\frac{1}{N_{\rm esc} (2\pi \sigma_v^2)^{3/2}}\exp{\left[ -\frac{({\bf v}+{\bf V}_{\rm lab})^2}{2 \sigma_v^2} \right]},
\label{VelDist}
\end{equation}
for $|{\bf v}+{\bf V}_{\rm lab}|<v_{\rm esc}$, and zero otherwise.
Here ${\bf v}$ is the WIMP velocity relative to the detector, and
\begin{equation}
N_{\rm esc}=\mathop{\rm erf}\left(\frac{v_{\rm esc}}{\sqrt{2}\sigma_v}\right)-\sqrt{\frac{2}{\pi}}\frac{v_{\rm esc}}{\sigma_v}\exp{\left[-\frac{v_{\rm esc}^2}{2\sigma_v^2} \right]}.
\label{Radon-transform}
\end{equation}
The laboratory is moving with velocity $\textbf{V}_{\rm lab}$ with respect to the Galaxy (thus  $- \textbf{V}_{\rm lab}$ is the average velocity of the WIMPs with respect to the detector). Following Ref.~\cite{Kuhlen},  which gives 100 km/s and 130 km/s as extreme estimates for the 1D velocity dispersion $\sigma_v/\sqrt{3}$, we take $\sigma_v$ either 173 km/s or 225 km/s. A recent study by the RAVE survey using high velocity stars finds an escape speed in the range $498~{\rm {km/s}}<v_{\rm esc}<608$~km/s at 90\% confidence, with a median-likelihood value of 544 km/s~\cite{RAVE}. We use the median RAVE value, $v_{\rm esc}=544$~km/s for most of our results, but for a few figures we keep  the older usual value $v_{\rm esc}=650$~km/s.
For the Galactic rotation speed $V_{\rm GalRot}$ at the position of the Sun we take 180 km/s and 312 km/s as low and high estimates  (see Appendix A).

Due to the ellipticity of the Earth's orbit the times of maximum and minimum $V_{\rm {lab}}$ are not exactly half a year apart. These times depend on $V_{\rm {Gal Rot}}$. For $V_{\rm {Gal Rot}}=180$ km/s, $V_{\rm {lab}}$ is maximum on May 30 and minimum on December 1. For $V_{\rm {Gal Rot}}=312$ km/s, the maximum and minimum happen on June 2 and December 5, respectively. Table~\ref{table:Vlab} gives the maximum and minimum values of $|{\bf V}_{\rm lab}|=V_{\rm {lab}}$ (in km/s) for our two choices of $V_{\rm GalRot}$, including the contribution of the Solar motion ${\bf V}_{\rm Solar}$.

\begin{table}[t]
\begin{center}
\begin{tabular}{lcccc}
\hline
\hline
Date & ~~~$V_{\rm {Gal Rot}}$ (km/s)~~~ & ~~~$\left|{\bf V}_{\rm {Gal Rot}}+{\bf V}_{\rm {Solar}}\right|$ (km/s)~~~ & $V_{\rm {lab}}$ (km/s)
\\
\hline
December 1 & 180 & 193 & 179.8 \\
May 30 & 180 & 193 & 208.8\\
\hline
December 5 & 312 & 324 & 310.5\\
June 2 & 312 & 324 & 340.2 \\
\hline
\hline
\end{tabular}
\caption{Maximum and minimum values of $V_{\rm lab}$ in km/s, for our two choices of the Galactic rotation speed $V_{\rm GalRot}$.}
\label{table:Vlab}
\end{center}
\end{table}

The Radon transform in the laboratory frame for the truncated Maxwellian WIMP velocity distribution in Eq.~\ref{VelDist}  is~\cite{Gondolo:2002}
\begin{equation}
\hat{f}_{\rm lab}\!\left( \frac{q}{2\mu}, \hat{\bf q} \right)=\frac{1}{{N_{\rm esc}(2\pi \sigma_v^2)^{1/2}}}~{\left\{\exp{\left[-\frac{\left[ (q/2\mu) + \hat{\bf q} \cdot {\bf V}_{\rm lab}\right]^2}{2\sigma_v^2}\right]}-\exp{\left[\frac{-v_{\rm esc}^2}{2\sigma_v^2}\right]}\right\}},
\label{fhatTM}
\end{equation}
if $(q/2\mu) + \hat{\bf q} \cdot {\bf V}_{\rm lab} < v_{\rm esc}$, and zero otherwise.

The recoil momentum ${\bf q}$ is measured in a reference frame fixed to the detector. The detector  frame is at some orientation in the laboratory frame, which we define as fixed to the Earth with axes pointing to the North, the West and the Zenith. The transformation equations  between the detector frame and the laboratory frame can be conveniently written in terms of  direction cosines measurable in any experiment. This is formulated in Appendix A, where we also give the transformations from the laboratory frame in any location on Earth to the Galactic reference frame.  The transformations in Appendix A take into account  Earth's rotation around its axis, which is usually neglected.

From Eq.~\ref{fhatTM} we can immediately see that if $q/2 \mu < |{\bf V}_{\rm lab}|$, i.e.\ $v_q < V_{\rm lab}$, the maximum of $\hat{ f}_{\rm lab} (\frac{q}{2 \mu},\hat{\textbf{q}})$ occurs when $-\hat{\bf q} \cdot {\bf V}_{\rm lab}=v_q$. Only if  $v_q > V_{\rm lab}$ the maximum is in the direction of $-{\bf V}_{\rm lab}$. Thus when the minimum WIMP speed $v_q$ required to cause a recoil momentum $q$ is less than $V_{\rm lab}$, the maximum of $\hat{f}_{\rm lab}$ occurs at an angle $\gamma$ between $\hat{\bf q}$ and  $-{\bf V}_{\rm lab}$ given by
\begin{equation}
\cos\gamma=\frac{v_q}{V_{\rm lab}}=\sqrt{\frac{M E_R}{2 \mu^2 V_{\rm lab}^2}}.
\label{gamma}
\end{equation}

\begin{figure}[t]
\begin{center}
  \includegraphics[height=200pt]{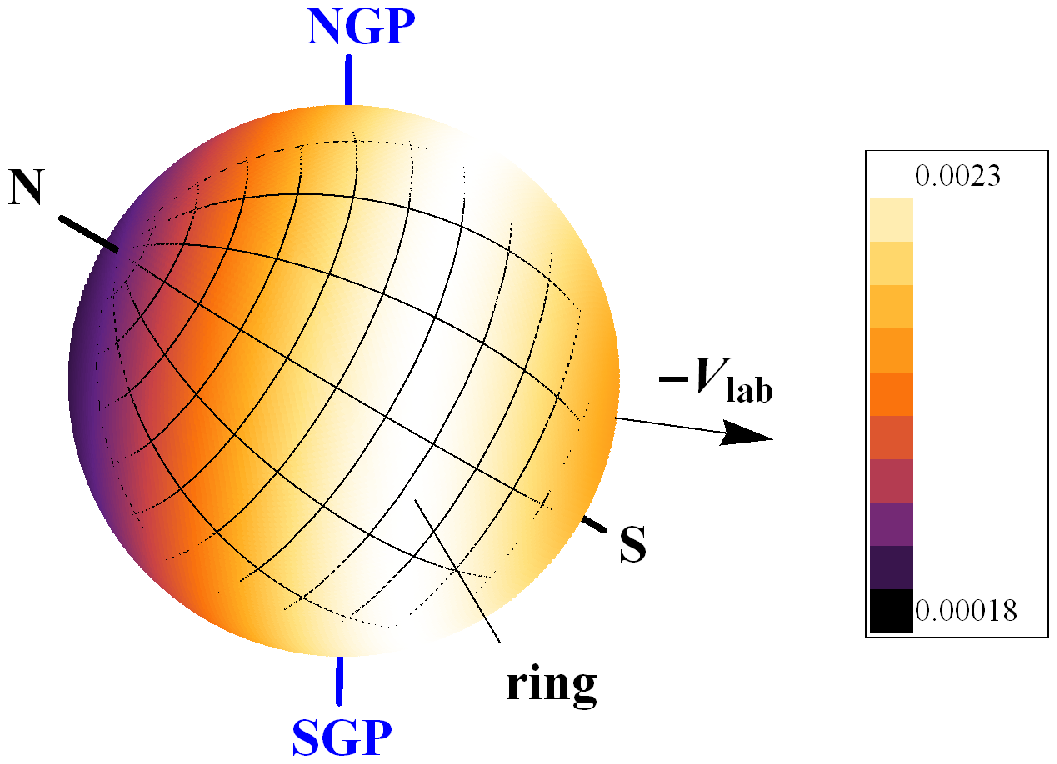}
  \includegraphics[height=190pt]{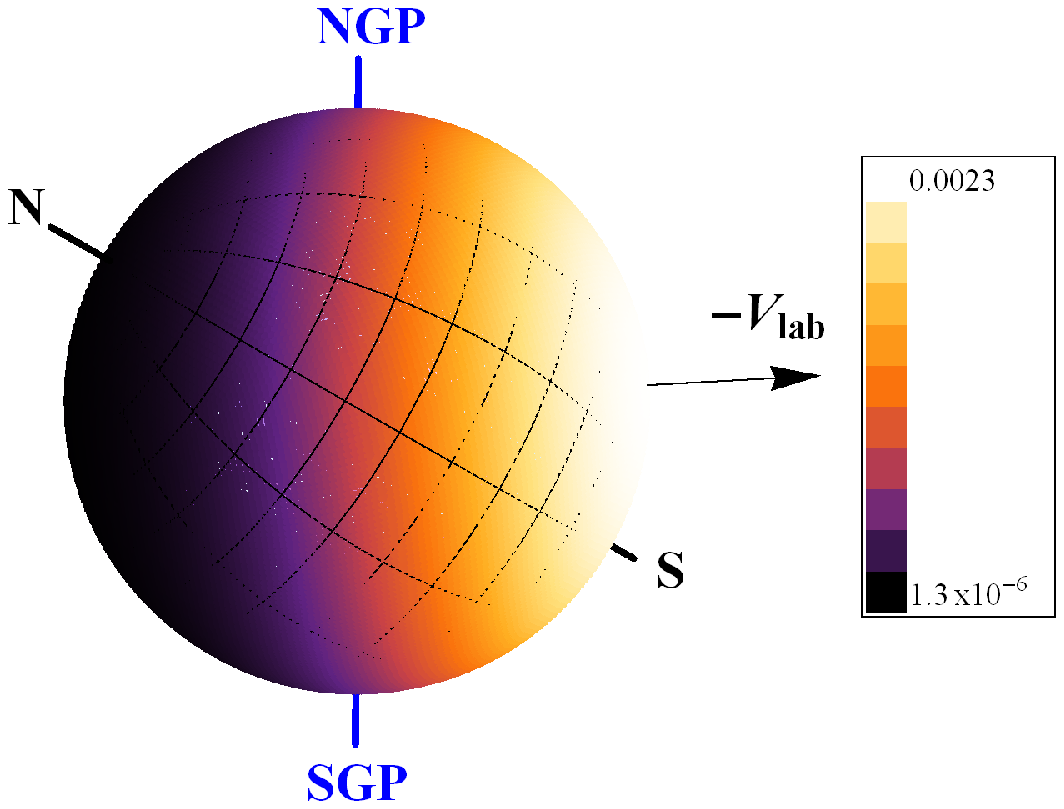}\\
  \vspace{-0.5cm}\caption{Angular recoil distribution $\hat{f}_{\rm lab}(v_q,\hat{\bf q})$ at fixed $v_q$ for sulfur recoils of 100 GeV/$c^2$ WIMPs on (a) December 4 with $E_R=5$ keV, thus $v_q=113$ km/s, and (b) June 2 with $E_R=40$ keV, thus $v_q=319$ km/s. The distribution is plotted on the sphere of recoil directions $\hat{\bf q}$ using the HEALPix pixelization. The IMB Model is used with parameters $V_{\rm {GalRot}}=280$ km/s, $\sigma_v=173$ km/s, and $v_{\rm esc}=650$ km/s. The North and South Galactic Poles (NGP and SGP), the North and South celestial poles (N and S), and the Earth meridians and parallels are indicated. The arrows show the direction of the average WIMP velocity  $-{\bf V}_{\rm lab}$, directed into the South Galactic Hemisphere (SGH) in December and into the North Galactic Hemisphere (NGH) in June. The color scale/grayscale shown in the vertical bars correspond to equal steps between the minimum and maximum values in units of ${(\text{km/s})}^{-1}$. Notice the ring of maximum $\hat{f}_{\rm lab}$ values in  (a) at $\gamma=66^\circ$ of $-{\bf V}_{\rm {lab}}$. }
  \label{Healpix-S}
\end{center}
\end{figure}

The ring is clearly visible when  $\hat{f}_{\rm lab}$ is plotted on the sphere  of recoil directions, as for example on panel (a) of Fig.~\ref{Healpix-S}. In this figure we use the HEALPix pixelization~\cite{HEALPix:2005} to plot $\hat{f}_{\rm lab}$ on the sphere of recoil directions for sulfur recoils at two different times in the year and recoil energies, assuming $V_{\rm GalRot}=280$ km/s, $\sigma_v=173$ km/s, $v_{\rm esc}=650$ km/s, and $m=100$ GeV/$c^2$. Fig.~\ref{Healpix-S}.a shows $\hat{f}_{\rm lab}$ on December 4 (when $V_{\rm lab}$ is minimum) for $E_R=5$ keV. We can see a ring of maximum $\hat{f}_{\rm lab}$, i.e.~of maximum recoil directional rate around $-{\bf V}_{\rm lab}$. $-{\bf V}_{\rm lab}$ points slightly towards the South Galactic Hemisphere. Fig.~\ref{Healpix-S}.b shows $\hat{f}_{\rm lab}$ on June 2 (when $V_{\rm lab}$ is maximum) and for $E_R=40$ keV. In this case the maximum of $\hat{f}_{\rm lab}$ and thus of the recoil rate, is in the direction of $-{\bf V}_{\rm lab}$. $-{\bf V}_{\rm lab}$ points slightly towards the North Galactic Hemisphere at this time. The color scale/grayscale plotted on the spheres in Fig.~\ref{Healpix-S} indicate different values  between 0.00018 ${(\text{km/s})}^{-1}$ and 0.0023 ${(\text{km/s})}^{-1}$ for Fig.~\ref{Healpix-S}.a and between $1.3 \times 10^{-6}$ ${(\text{km/s})}^{-1}$ and 0.0023 ${(\text{km/s})}^{-1}$ for Fig.~\ref{Healpix-S}.b as shown in the horizontal bar. The minimum WIMP speed is $v_q=113$ km/s and $V_{\rm lab}=278.7$ km/s in Fig.~\ref{Healpix-S}.a and $v_q=319$ km/s and $V_{\rm lab}=308.3$ km/s in Fig.~\ref{Healpix-S}.b.

  \begin{figure}[t]
\begin{center}
  \includegraphics[height=170pt]{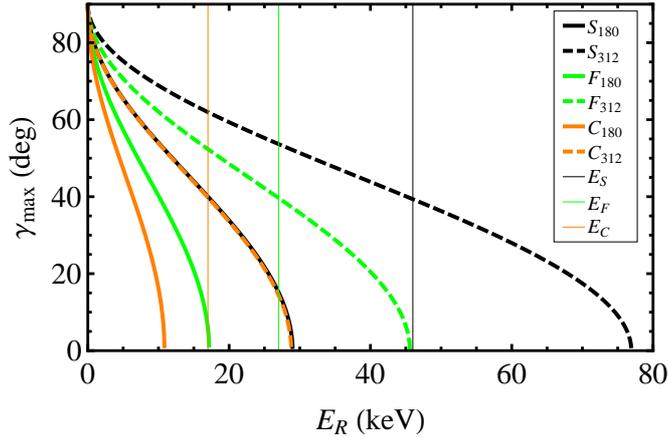}\\
  \vspace{-0.1cm}\caption{Maximum ring radius $\gamma_{\rm {max}}$, measured in degrees from the direction of $-{\bf V}_{\rm lab}$, as a function of $E_R$, for three target elements S, F, and C. The labels indicate the target element, the subscripts showing the value of $V_{\rm GalRot}$ in km/s. The sulfur solid  black line S$_{180}$ coincides with the carbon dashed orange line C$_{312}$. The possible values of the ring radius $\gamma$ lie below the respective curves, and to the left of the vertical lines (labeled $E_S$, $E_F$, and $E_C$ for S, F, and C, respectively), where $\hat{f}_{\rm center}/\hat{f}_{\rm ring}<0.9$ (see Eq.~\ref{fcm}). The ring would be very difficult to observe with a smaller contrast, namely for larger $\hat{f}_{\rm center}/\hat{f}_{\rm ring}$.}
  \label{cosgamma}
\end{center}
\end{figure}

From Eq.~\ref{gamma}, the ring of maximum $\hat{f}_{\rm lab}$ exists if $E_R<2\mu^2 V_{\rm lab}^2/M$. For example, for WIMPs of mass $m=100$ GeV/$c^2$ and  $V_{\rm lab}=278.7$ km/s, as in Fig.~\ref{Healpix-S}.a, a ring exists in S recoils if $E_R < 30.6$ keV. Since the reduced mass approaches its maximal value $\mu=M$ as $m\gg M$, there can be no ring if $E_R>2M  V_{\rm lab}^2$ for any value of $m$ and the maximum possible ring radius (and thus the minimum possible $\cos \gamma$) is given by
\begin{equation}
\cos \gamma_{\rm {max}}=\sqrt{\frac{E_R}{2 M V_{\rm lab}^2}}.
\label{maxgamma}
\end{equation}
For $E_R=2M  V_{\rm lab}^2$, $\gamma_{\rm {max}}$ becomes zero. Fig.~\ref{cosgamma} shows plots of $\gamma_{\rm {max}}$  as a function of $E_R$ when $V_{\rm lab}$ is maximum for different target elements relevant to directional detectors. Here, either $V_{\rm GalRot}=180$ km/s or 312 km/s. The allowed values of $\gamma$ are below the lines shown.

The ring-like recoil feature we discuss here depends only on $\hat{f}_{\rm lab}$. In particular, it is independent of the type of WIMP interaction. However, to give concrete examples of how the ring would appear in an actual recoil rate, we show the ring for WIMPs with spin-independent interaction and equal couplings to protons and neutrons in a CS$_2$ detector. For these WIMPs, the cross section in Eq.~\ref{eq: rate} is
 $\sigma_i^{\rm SI}(q)=\mu_i^2 A_i^2 \sigma_p S_i(q)/\mu_p^2$, where $\mu_p=m m_p/(m+m_p)$ is the WIMP-proton reduced mass, $\sigma_p$ is the WIMP-proton cross section, $A_i$ is the mass number of the nuclear species $i$, and $S_i(q)$ is the nuclear form factor, for which we use the Helm~\cite{Helm:1956} nuclear form factor normalized to 1. The SI directional recoil rate is therefore~\cite{Alenazi-Gondolo:2008}
\begin{equation}
\frac{dR}{dE_R~d\Omega_q} = 1.306\times10^{-3}\frac{\text{events}}{\text{kg-day-keV-sr}}\times \frac{\rho_{0.3}~ \sigma_{44}}{4\pi m \mu_p^2} \sum_i C_i \, A_i^2 \,  \, S_i(q) \hat{f}_{\rm lab}\!\left( \frac{q}{2\mu_i}, \hat{\bf q} \right).
\label{RecoilRate}
\end{equation}
Here $\rho_{0.3}$ is the dark matter density in units of 0.3 GeV/$c^2$/cm$^3$, $\sigma_{44}$ is the WIMP-proton cross section in units of $10^{-44}\;{\text{cm}}^2$, $\mu_p$ and $m$ are in GeV/$c^2$, and $\hat{ f}_{\rm lab}$ is in ${(\text{km/s})}^{-1}$.
The standard value of the local dark matter density is $\rho=0.3$ GeV/$c^2$/cm$^3$, and this is what we use here. However, one should keep in mind the  large uncertainties in this parameter. Recent astronomical constraints are consistent with 0.2 GeV/$c^2$/cm$^3$ $<\rho<0.4$ GeV/$c^2$/cm$^3$ for a spherical dark matter halo profile, and up to 20\% larger for non-spherical haloes~\cite{Weber:2010}. Using a halo model independent method, Ref.~\cite{Salucci:2010} finds $\rho=0.43 \pm 0.11 \pm 0.10$ GeV/$c^2$/cm$^3$, with uncertainties from two different sources. Ref.~\cite{Pato:2010} finds the density for a specific simulated galaxy resembling the Milky Way is 21\% larger than the mean value of $\rho=0.39$ GeV/$c^2$/cm$^3$ obtained in a previous study~\cite{Catena:2010} in which spherical symmetry was assumed.

\begin{figure}[t]
\begin{center}
  \includegraphics[height=200pt]{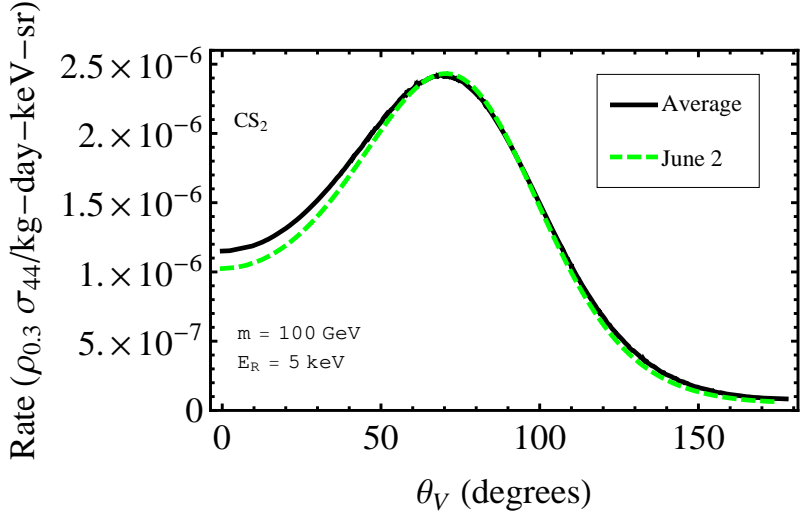}
  \vspace{-0.1cm}\caption{Directional differential recoil rate as a function of the polar angle $\theta_V$ measured from $-{\bf V}_{\rm lab}$  for $V_{\rm GalRot}=312$ km/s, $\sigma_v=173$ km/s, $v_{\rm esc}=544$ km/s, $m=100$ GeV/$c^2$, $E_R=5$ keV in CS$_2$. The solid black curve is the average rate over the year 2011, and the dashed green curve is the rate on June 2, when $V_{\rm lab}$ is maximum.}
 \label{TimeAvgRate}
\end{center}
\end{figure}

Fig.~\ref{Healpix-S}.a shows the ring at a particular time. However, in direct dark matter searches the event rate will need to be measured during several years to obtain the required exposure to observe the ring. Fig.~\ref{TimeAvgRate} shows that the recoil rate averaged over a year is very similar to the distribution computed at a particular time. The figure shows the plot of the directional rate in CS$_2$ averaged over one year (solid black curve) and on June 2 (dashed green curve) as a function of the opening angle $\theta_V$ between the recoil direction and the average $-{\bf V}_{\rm lab}$ over the year, at one fixed azimuthal angle $\phi_V$. In the IMB, for a given $\theta_V$ the rate is the same at different azimuthal angles around $-{\bf V}_{\rm lab}$. Here $V_{\rm GalRot}=312$ km/s, $\sigma_v=173$ km/s, $v_{\rm esc}=544$ km/s, $m=100$ GeV/$c^2$, and $E_R=5$ keV. If a ring exists, the rate reaches its maximum at $\theta_V=\gamma$. The two curves shown in Fig.~\ref{TimeAvgRate} are very similar, thus the effect of integrating the rate over time is not significant. For simplicity, in this paper we compute the rate at specific dates only (we choose those at which $V_{\rm lab}$ is maximum).

\begin{figure}[t]
\begin{center}
  \includegraphics[height=90pt]{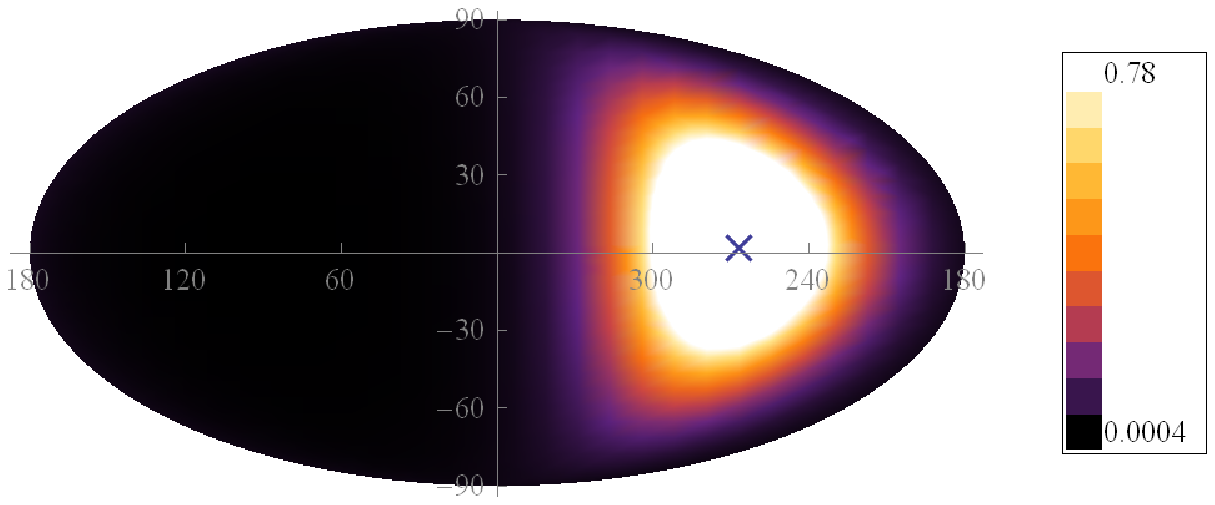}
  \includegraphics[height=90pt]{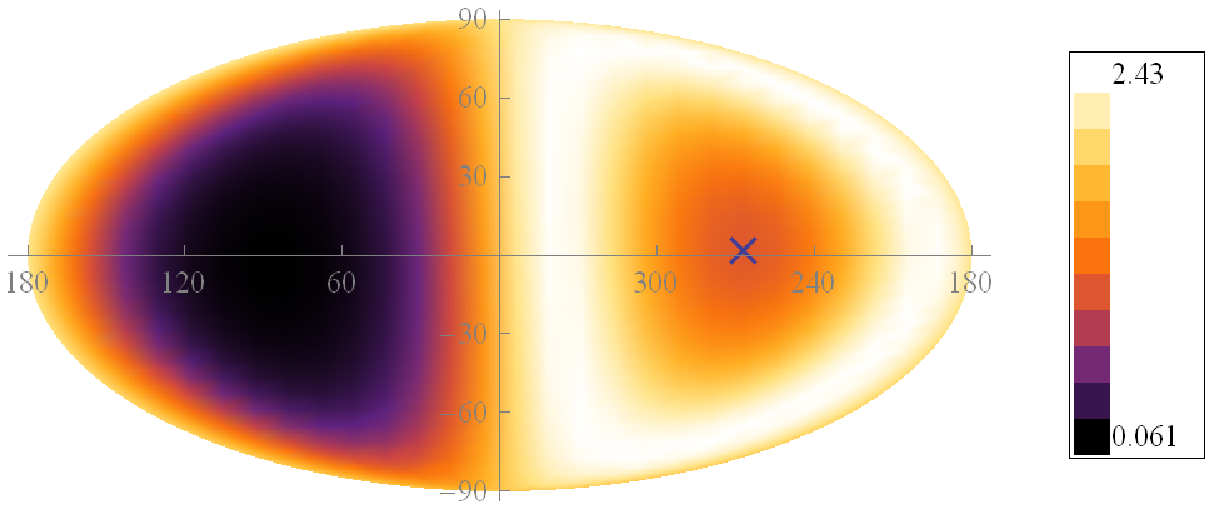}\\
  \vspace{-0.1cm}\caption{Mollweide equal-area projection maps of the celestial sphere in Galactic coordinates showing (a) the number fraction  $F_{\rm WIMP}(\hat{\bf v},v_q)$ of $m = 100$ GeV/$c^2$  WIMPs crossing the Earth  per unit solid angle as a function of the WIMP velocity direction $\hat{\bf v}$. For this figure we take a minimum speed $v_q=113$ km/s, as necessary to produce $E_R=5$ keV sulfur recoils. (b) The directional differential recoil rate in CS$_2$ at $E_R=5$ keV for $m = 100$ GeV/$c^2$. In both panels we assume the IMB with $v_{\rm esc}=544$ km/s, $\sigma_v=173$ km/s and $V_{\rm GalRot}=312$ km/s on June 2. Notice the direction of $-{\bf V}_{\rm lab}$ marked with  a cross. The color scale/grayscale shown in the vertical bars corresponds to  equal steps between the minimum and maximum values in \protect\ref{Flux}.a in units of sr$^{-1}$, and in \protect\ref{Flux}.b in units of $10^{-6} \times (\rho_{0.3} \sigma_{44}/{\text{kg-day-keV-sr}})$. Eq.~\ref{fcm} gives $\hat{f}_{\rm center}/\hat{f}_{\rm ring}=0.42$ in the right panel.}
  \label{Flux}
\end{center}
\end{figure}

\begin{figure}[t]
\begin{center}
  \includegraphics[height=90pt]{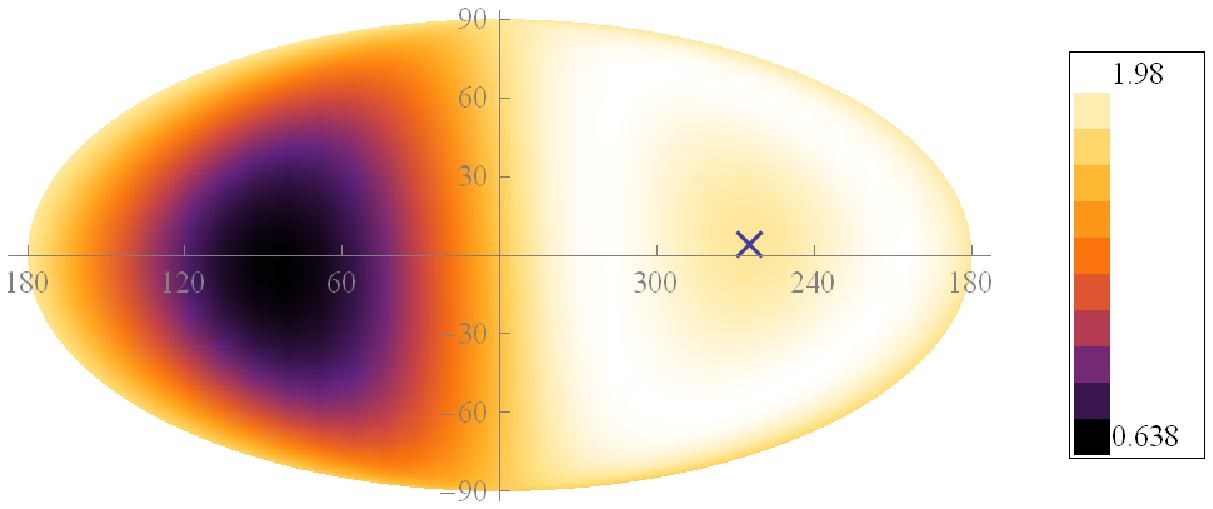}
  \includegraphics[height=90pt]{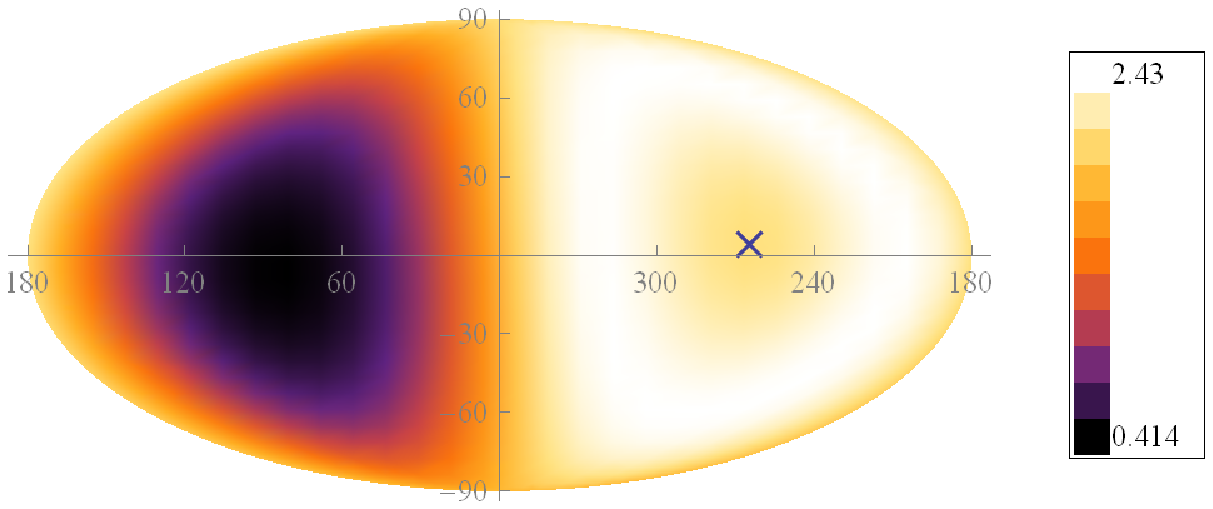}\\
  \includegraphics[height=90pt]{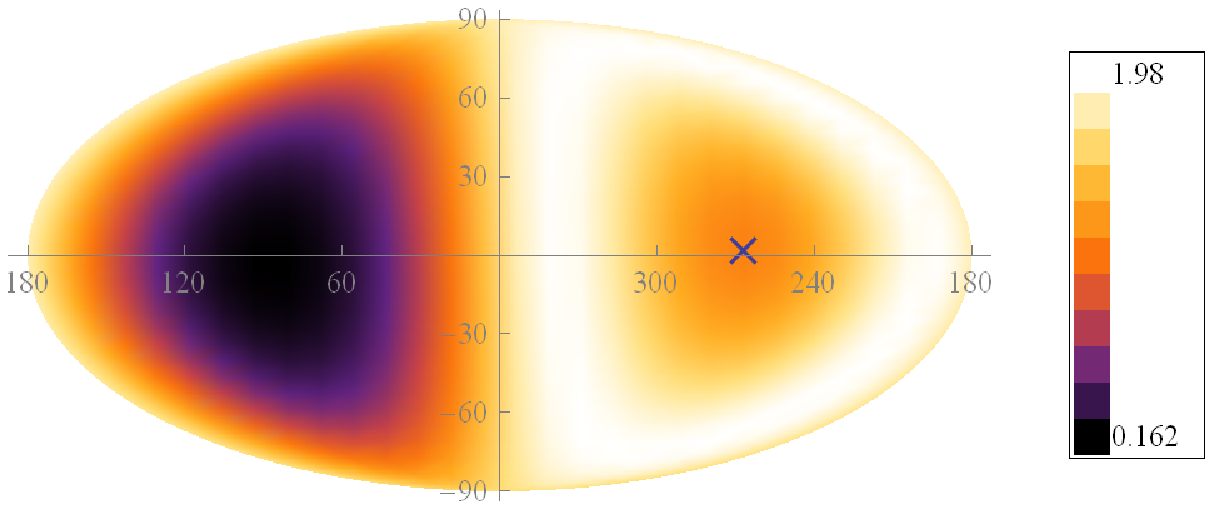}\\
  \vspace{-0.1cm}\caption{Directional differential recoil rate for different values of the Galactic rotation velocity $V_{\rm GalRot}$ and of the WIMP velocity dispersion $\sigma_v$:
  (a) $V_{\rm GalRot}=180$ km/s,  $\sigma_v=225$ km/s on May 30;
  (b) $V_{\rm GalRot}=180$ km/s,  $\sigma_v=173$ km/s on May 30;
  (c) $V_{\rm GalRot}=312$ km/s,  $\sigma_v=225$ km/s on June 2.
The right panel of Fig.~\ref{Flux} displays the fourth combination $V_{\rm GalRot}=312$ kms/ and $\sigma_v=173$ km/s.  All panels have $E_R=5$ keV in CS$_2$,  $m=100$ GeV/$c^2$, and $v_{\rm esc}=544$ km/s (the same parameters as in the right panel of Fig.~\ref{Flux}). The color scale/grayscale shown in the vertical bars correspond to equal steps between the minimum and maximum values given in units of $10^{-6} \times (\rho_{0.3} \sigma_{44}/{\text{kg-day-keV-sr}})$. In panels (a), (b), and (c), $\hat{f}_{\rm center}/\hat{f}_{\rm ring}=0.91$, 0.86, and 0.60, respectively.}
 \label{VGalRot-sigmav}
\end{center}
\end{figure}

The plots of $\hat{f}_{\rm lab}$ or directional rate are better viewed not in a HEALPix sphere, but in a planar projection. In Fig.~\ref{Flux} and \ref{VGalRot-sigmav} we use the Mollweide equal-area projection maps of the celestial sphere in Galactic coordinates. The relationship between the ($x$,$y$) coordinates on the Mollweide map and the Galactic longitude and latitude ($l$,$b$) is given by~\cite{Gelmini-Gondolo:2001}
\begin{equation}
l=\frac{-\pi x}{2\sqrt{2}\cos\theta},
\qquad
b=\arcsin{\left(\frac{2\theta+\sin(2\theta)}{\pi}\right)},
\end{equation}
where
\begin{equation}
\theta=\arcsin{(y/\sqrt{2})}.
\end{equation}

We first show, in Fig.~\ref{Flux}.a, a Mollweide projection of the number fraction of WIMPs $F_{\rm WIMP}(\hat{\bf v},v_q)$ crossing the detector per unit solid angle with speed larger than ${v_q}$, as a function of the WIMP velocity direction $\hat{\bf v}$ (these WIMPs produce a recoil momentum of magnitude $q$ or higher when scattering off a nucleus of mass $M$)~\cite{ChanIV}
\begin{equation}
F_{\rm WIMP}(\hat{\bf v},v_q)=\int_{v_q}^{v_{\rm max}(\hat{\bf v})}{f_{\rm WIMP}({\bf v}) v^2 dv}.
\end{equation}
The upper limit of this integral is
$v_{\rm max}(\hat{\bf v})=-\hat{\bf v} \cdot {\bf V}_{\rm lab}+\sqrt{(\hat{\bf v} \cdot {\bf V}_{\rm lab})^2-{\bf V}_{\rm lab}^2+v_{\rm esc}^2}$
and the analytic expression of  $F_{\rm WIMP}(\hat{\bf v},v_q)$ is given in Eq.~13 of Ref.~\cite{ChanIV}. The maximum of $F_{\rm WIMP}(\hat{\bf v},v_q)$ happens when $\hat{\bf v} \cdot {\bf V}_{\rm lab}=-V_{\rm lab}$, i.e. in the direction of the  average WIMP velocity $-{\bf V}_{\rm lab}$. Most WIMPs move in the direction opposite to the laboratory motion, marked by a cross in the figures.

In Fig.~\ref{Flux}.b we show a Mollweide map of the directional differential recoil rate in CS$_2$, Eq.~\ref{RecoilRate}, produced by the WIMPs in Fig.~\ref{Flux}.a in which the ring of maximum rate around the $-{\bf V}_{\rm lab}$ direction is clearly visible. In Fig.~\ref{Flux}, we used $m = 100$ GeV/$c^2$, and  the IMB with $v_{\rm esc}=544$ km/s, $V_{\rm GalRot}=312$ km/s and $\sigma_v=173$ km/s on June 2. In Fig.~\ref{Flux}.b the recoil energy is $E_R=5$ keV.

\begin{figure}[t]
\begin{center}
  \includegraphics[height=130pt]{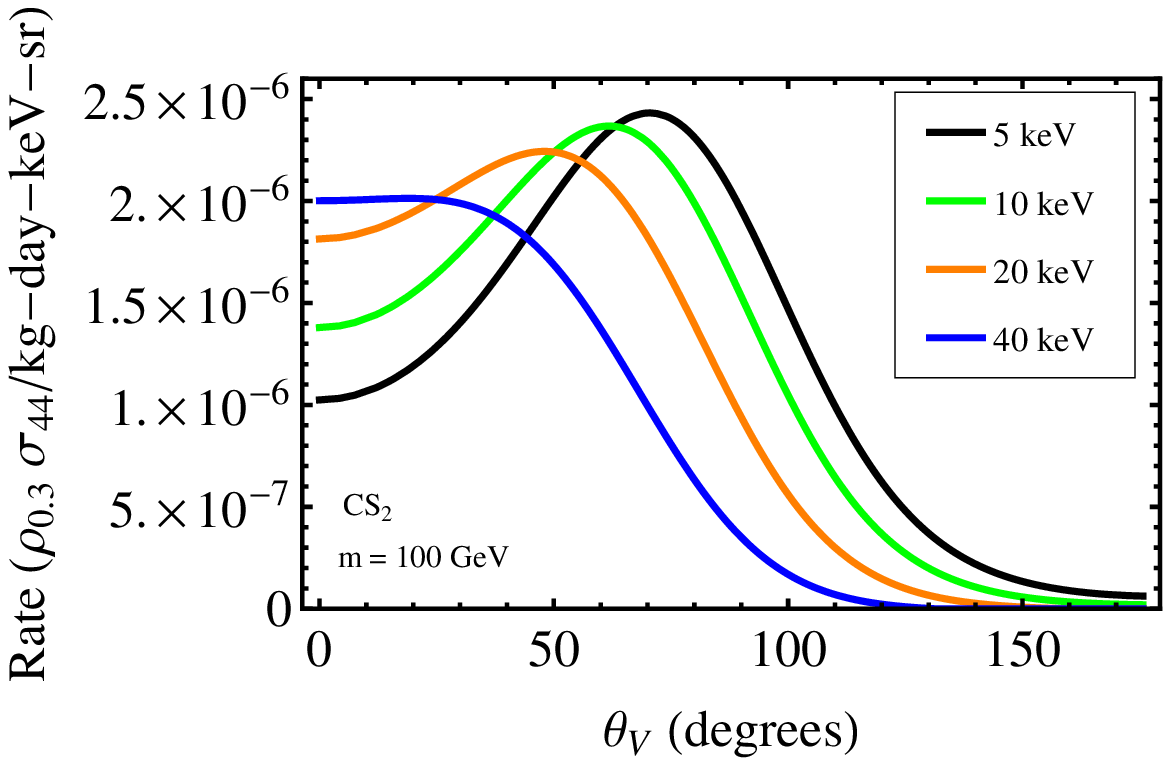}
  \includegraphics[height=130pt]{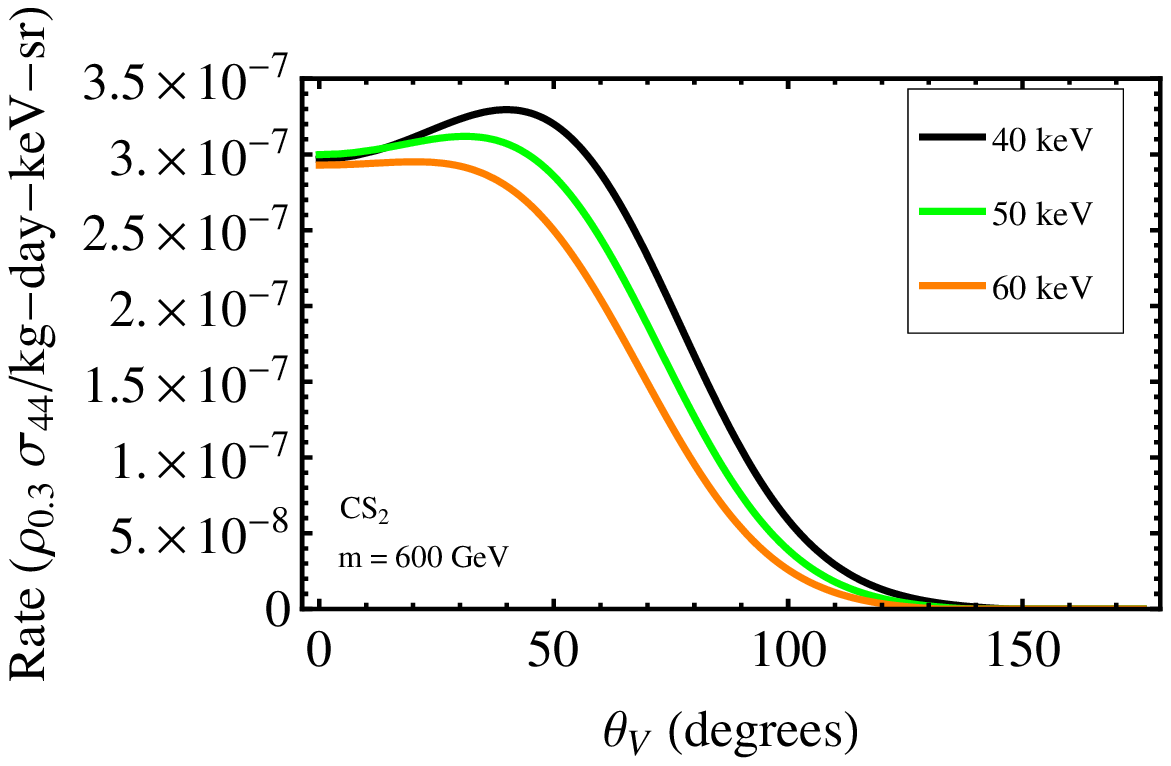}\\
  \vspace{-0.1cm}\caption{Directional differential recoil rate as a function of the polar angle $\theta_V$ measured from $-{\bf V}_{\rm lab}$
  for  CS$_2$, with $V_{\rm GalRot}=312$ km/s,  $\sigma_v=173$ km/s on June 2, for (a) $m = 100$ GeV/$c^2$ at $E_R=5$, 10, 20, and 40 keV (Fig.~\ref{Flux}.b corresponds to the 5 keV profile) and (b) $m = 600$ GeV/$c^2$ at $E_R=40$, 50, and 60 keV.}
  \label{RateAngle-CS2}
\end{center}
\end{figure}

It is easier to see the ring when the contrast between the rate at the center of the ring (in the direction of $-{\bf V}_{\rm lab}$) and the ring is larger. In terms of $\hat{f}_{\rm lab}$, the ratio of the value $\hat{f}_{\rm center}$ at the center of the ring to the value $\hat{f}_{\rm ring}$ at the ring is approximately, for the IMB neglecting the escape speed,
\begin{equation}
\frac{\hat{f}_{\rm center}}{\hat{f}_{\rm ring}} \simeq \exp{\left[-\frac{(V_{\rm lab}-v_q)^2}{2 \sigma_v^2} \right]}.
\label{fcm}
\end{equation}
$\hat{f}_{\rm center}/\hat{f}_{\rm ring}=0.42$ in Fig.~\ref{Flux}.b (see the 5 keV profile in  Fig.~\ref{RateAngle-CS2}.a).
The smaller the ratio $\hat{f}_{\rm center}/\hat{f}_{\rm ring}$, the easier it is to detect the ring. Thus the best prospects to observe the ring are at low recoil energies and for heavier WIMPs (so $v_q$ is small), large $V_{\rm lab}$ and small $\sigma_v$. In Figs.~\ref{Flux}.b and \ref{VGalRot-sigmav} we vary $V_{\rm lab}$ and $\sigma_v$ giving the four combinations of maximum and minimum values for both. Fig.~\ref{VGalRot-sigmav} shows plots of the CS$_2$ directional rate for $E_R=5$ keV, $m=100$ GeV/$c^2$, and three combinations of $V_{\rm lab}$ and $\sigma_v$ different from those in Fig.~\ref{Flux}: (a) $V_{\rm GalRot}=180$ km/s, $\sigma_v=225$ km/s on May 30; (b) $V_{\rm GalRot}=180$ km/s, $\sigma_v=173$ km/s on May 30; and (c) $V_{\rm GalRot}=312$ km/s, $\sigma_v=225$ km/s on June 2. The right panel of Fig.~\ref{Flux} displays the fourth combination of $V_{\rm GalRot}$ and $\sigma_v$.
The rate is dominated by scattering off S, by a factor of about 100. In Fig.~\ref{VGalRot-sigmav}.a, $V_{\rm lab}=208.8$ km/s and $\sigma_v=225$ km/s is the worst combination of low $V_{\rm lab}$ and high $\sigma_v$. It is clearly seen from the figures that the ring is most visible in Fig.~\ref{Flux}.b, with the largest $V_{\rm lab}$ and smallest $\sigma_v$ combination of the four. This best combination is chosen for Fig.~\ref{RateAngle-CS2}.

The dependence of the ratio $\hat{f}_{\rm center}/\hat{f}_{\rm ring}$ on energy can be extracted from Fig.~\ref{RateAngle-CS2}. Fig.~\ref{RateAngle-CS2} shows the CS$_2$ directional differential rate as a function of $\theta_V$ at different recoil energies for $V_{\rm lab}=312$ km/s and $\sigma_v=173$ km/s on June 2. In Fig.~\ref{RateAngle-CS2}.a, $m=100$ GeV/$c^2$, and the ring is present at all energies below 40 keV. Fig.~\ref{RateAngle-CS2}.b shows the case of $m=600$ GeV/$c^2$ at higher energies of $E_R=40$ keV, 50 keV and 60 keV. The ring is present at all energies below 50 keV in this case. In Fig.~\ref{RateAngle-CS2}.a, $\hat{f}_{\rm center}/\hat{f}_{\rm ring}=0.42$, 0.58, 0.80, and 0.99 for $E_R=5$, 10, 20, and 40 keV, respectively. In Fig.~\ref{RateAngle-CS2}.b, $\hat{f}_{\rm center}/\hat{f}_{\rm ring}=0.89$, 0.96, and 0.99 for $E_R=40$, 50, and 60 keV, respectively. For the same energy, $v_q$ is smaller for a larger $m$. Thus as $m$ increases, the ratio $\hat{f}_{\rm center}/\hat{f}_{\rm ring}$ becomes smaller and the contrast between the center and the ring becomes larger. This is the reason that the ring is present for $E_R=40$ keV in Fig.~\ref{RateAngle-CS2}.b but not in Fig.~\ref{RateAngle-CS2}.a.

\begin{figure}[t]
\begin{center}
  \includegraphics[height=130pt]{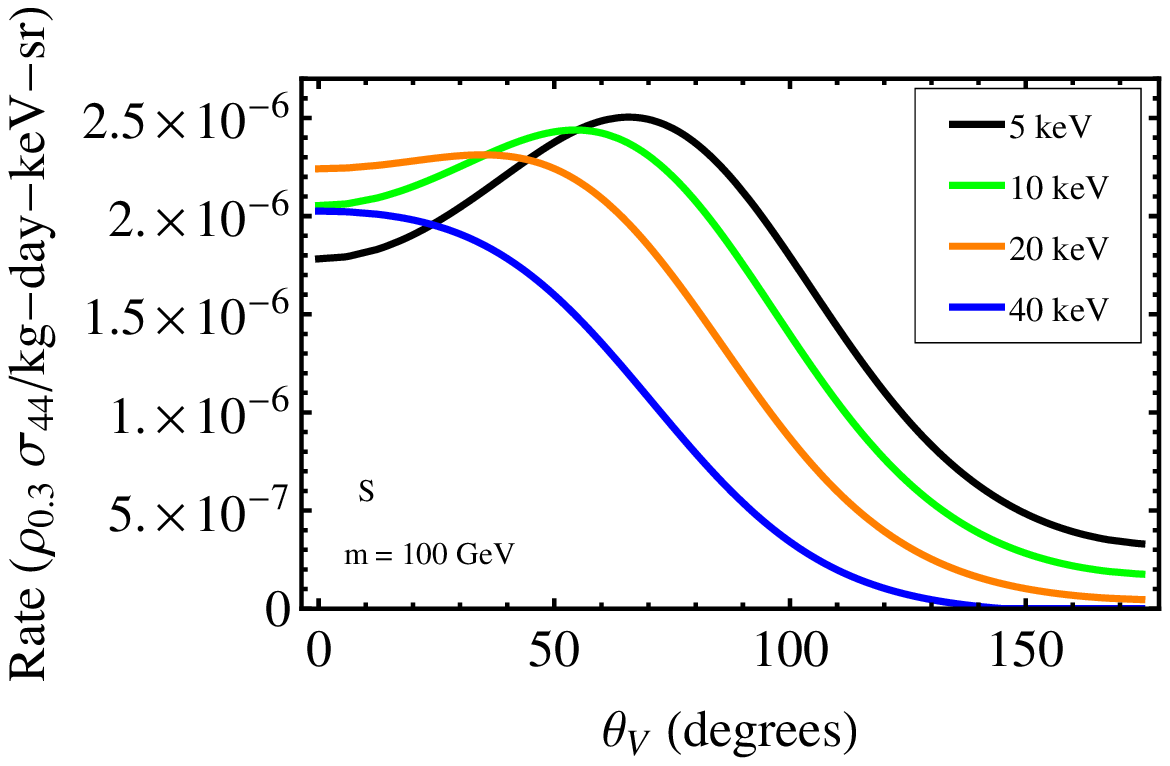}
  \includegraphics[height=130pt]{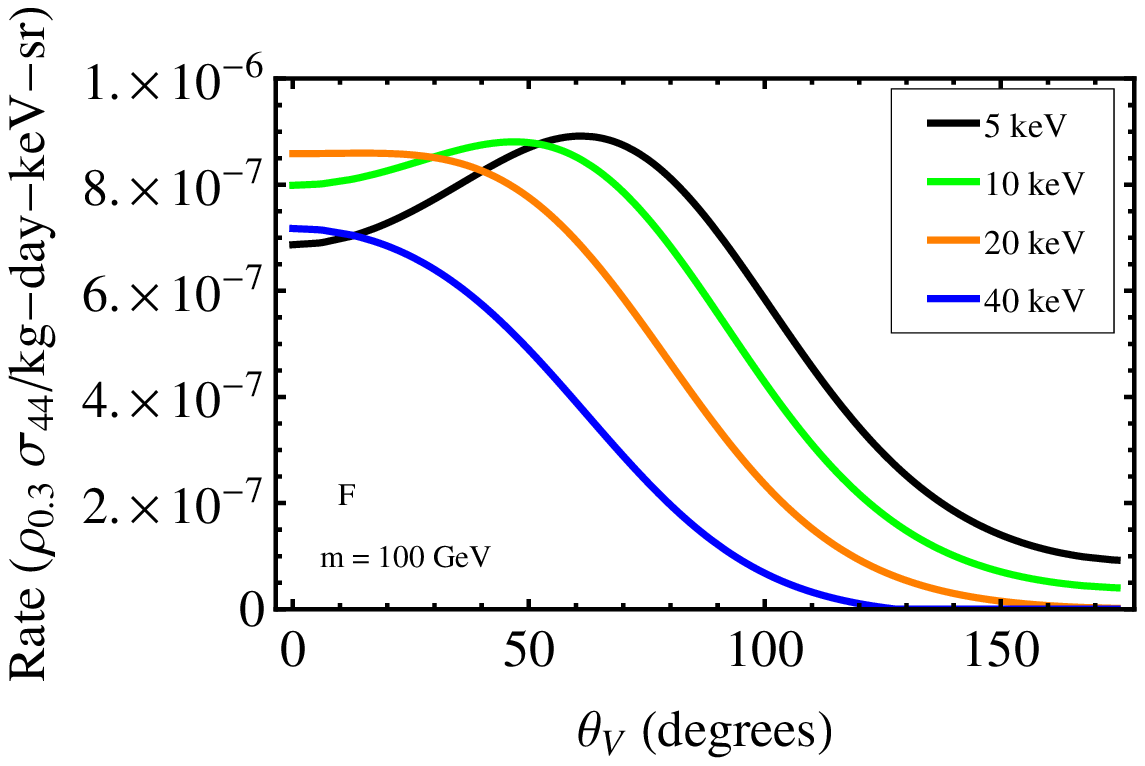}\\
  \includegraphics[height=130pt]{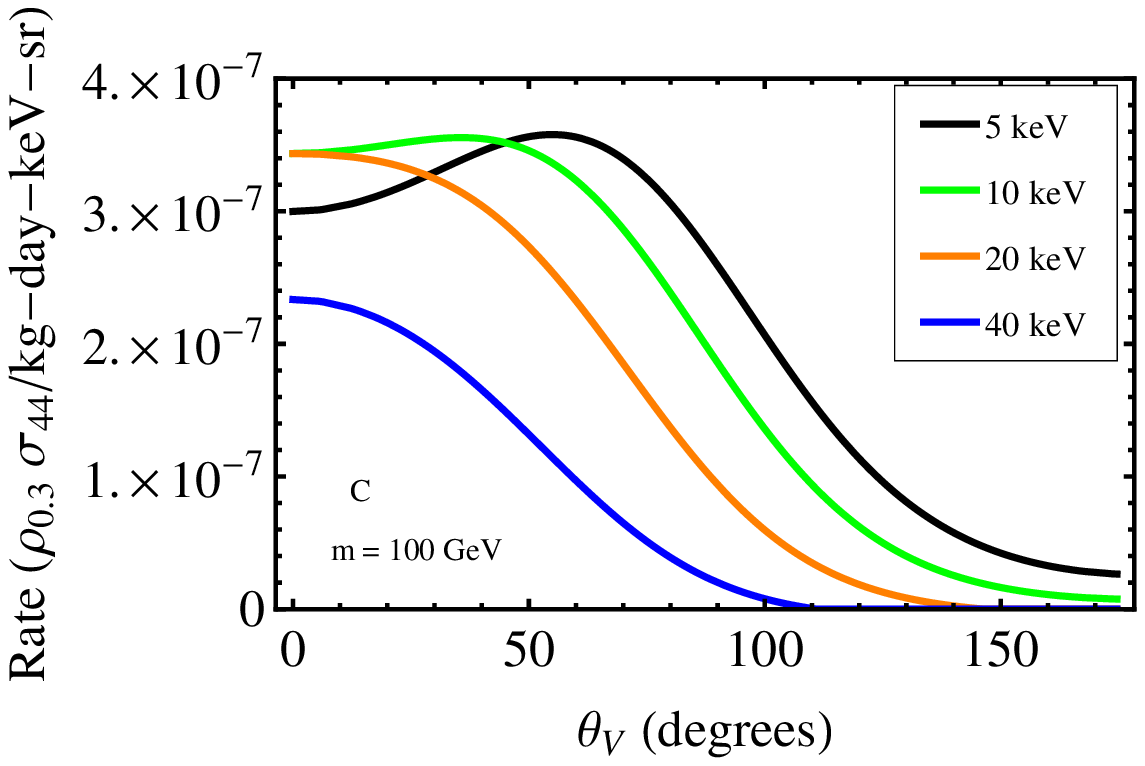}\\
  \vspace{-0.1cm}\caption{Same as Fig.~\ref{RateAngle-CS2}.a but  for (a) S, (b) F, and (c) C recoils in the IMB with  $v_{\rm esc}=544$ km/s, $\sigma_v=199$ km/s and $V_{\rm GalRot}=246$ km/s on June 1. Four recoil energies are shown in each plot: 5, 10, 20, and 40 keV.}
  \label{RateAngle-SFC}
\end{center}
\end{figure}

For other materials the ring's dependence on energy is explored in Fig.~\ref{RateAngle-SFC}, which shows the directional differential recoil rate for S, F, and C as a function of $\theta_V$ at different recoil energies, 5, 10, 20, and 40 keV. In this figure again $m=100$ GeV/$c^2$, but we use the average values of $\sigma_v$ and $V_{\rm lab}$, $\sigma_v=199$ km/s and $V_{\rm GalRot}=246$ km/s, which yields the maximum $V_{\rm lab}=274.4$ km/s on June 1, instead of the most favorable combination of values used in Fig.~\ref{RateAngle-CS2}. Notice that since in CS$_2$ the rate is dominated by scattering off S,  Figs.~\ref{RateAngle-CS2}.a and \ref{RateAngle-SFC}.a only differ because  $V_{\rm lab}$ is larger and $\sigma_v$ is smaller, thus for energies below 40 keV, the ratio $\hat{f}_{\rm center}/\hat{f}_{\rm ring}$ is smaller and the contrast between the ring and the center is larger in Fig.~\ref{RateAngle-CS2}.a.

In Fig.~\ref{RateAngle-SFC}.a, $\hat{f}_{\rm center}/\hat{f}_{\rm ring}=0.72$, 0.85, and 0.97 for $E_R=5$, 10, 20 keV, respectively. In Fig.~\ref{RateAngle-SFC}.b,  $\hat{f}_{\rm center}/\hat{f}_{\rm ring}=0.78$, 0.91, and $\simeq 1$ for $E_R=5$, 10, and 20 keV, respectively. In both Figs.~\ref{RateAngle-SFC}.a and \ref{RateAngle-SFC}.b, $v_q>V_{\rm lab}$ for $E_R=40$ keV. In Fig.~\ref{RateAngle-SFC}.c, $\hat{f}_{\rm center}/\hat{f}_{\rm ring}=0.84$ and 0.97 for $E_R=5$ and 10 keV, respectively, and $v_q>V_{\rm lab}$ for $E_R=20$ and 40 keV.

Fig.~\ref{RateAngle-SFC} illustrates the impact of using different targets on the existence of the ring in. When $m \gg M$, which is the case for all three panels in Fig.~\ref{RateAngle-SFC}, $v_q \simeq \sqrt{E_R/2M}$, and as the target mass $M$ increases, $v_q$ decreases and the contrast between the ring and center becomes larger. In Fig.~\ref{RateAngle-SFC}.a and Fig.~\ref{RateAngle-SFC}.b, the ring is present at energies below  20 keV, whereas in Fig.~\ref{RateAngle-SFC}.c, because C is lighter than S and F, the ring is present only at energies below 10 keV.

Measuring the differential rate will require very large statistics. Thus we explore next the probability of observing the ring in the rate integrated over energy.

\begin{figure}[t]
\begin{center}
  \includegraphics[height=130pt]{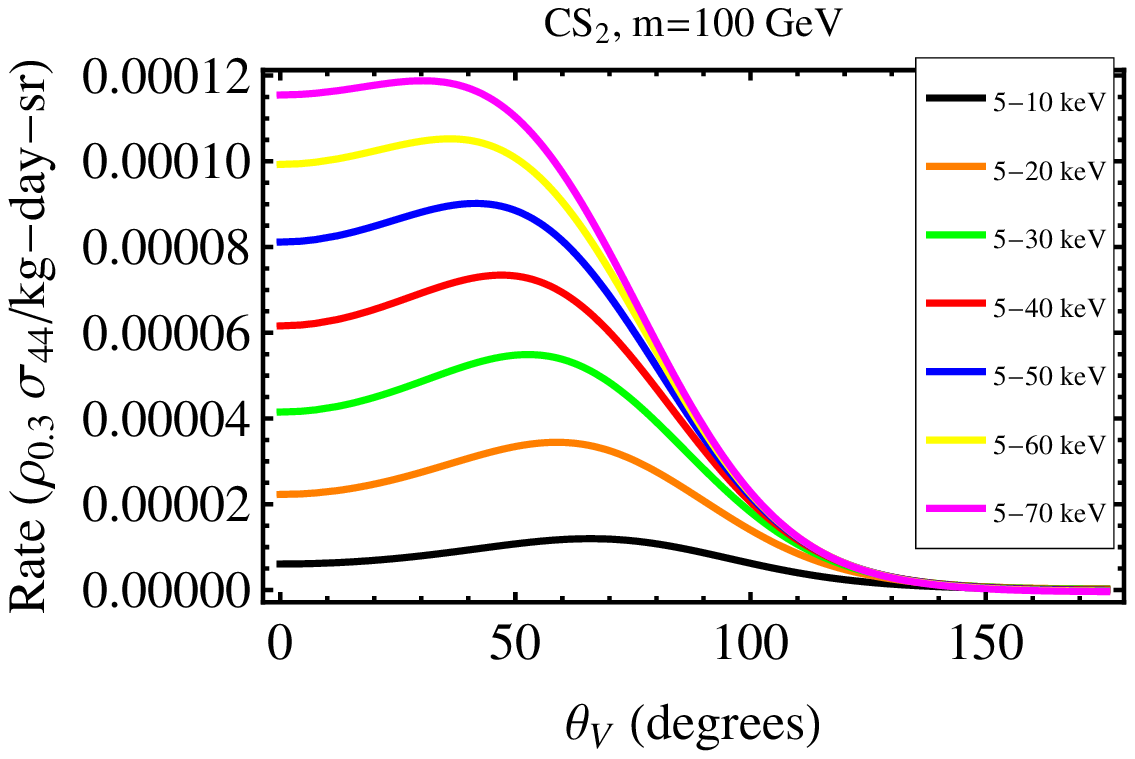}
  \includegraphics[height=130pt]{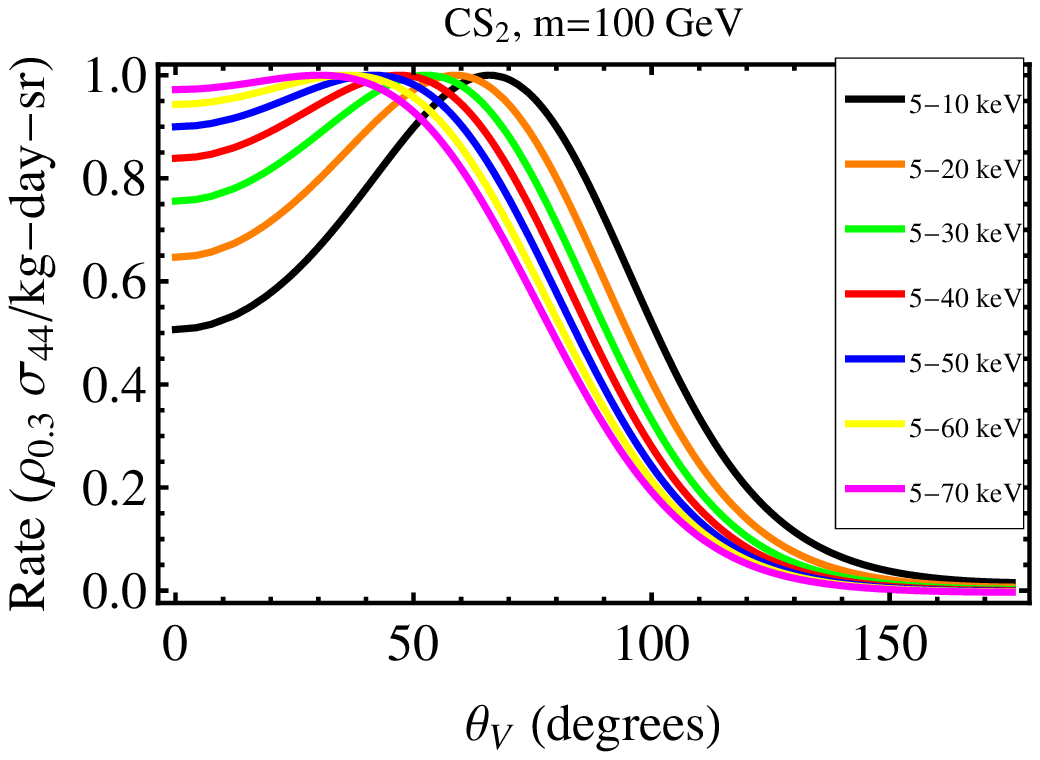}\\
  \vspace{-0.1cm}\caption{(a) Energy-integrated directional rate $\mathcal{R}(E_1,E_2,\theta_V)$ (Eq.~\ref{EnergyIntRate}) in  CS$_2$ for different energy intervals as a function of $\theta_V$ using the same parameters of Fig.~\ref{RateAngle-CS2}. (b) Same rates re-scaled to be between 0 and 1.}
  \label{RateAngle-CS2-Integrated}
\end{center}
\end{figure}

\begin{figure}[t]
\begin{center}
  \includegraphics[height=130pt]{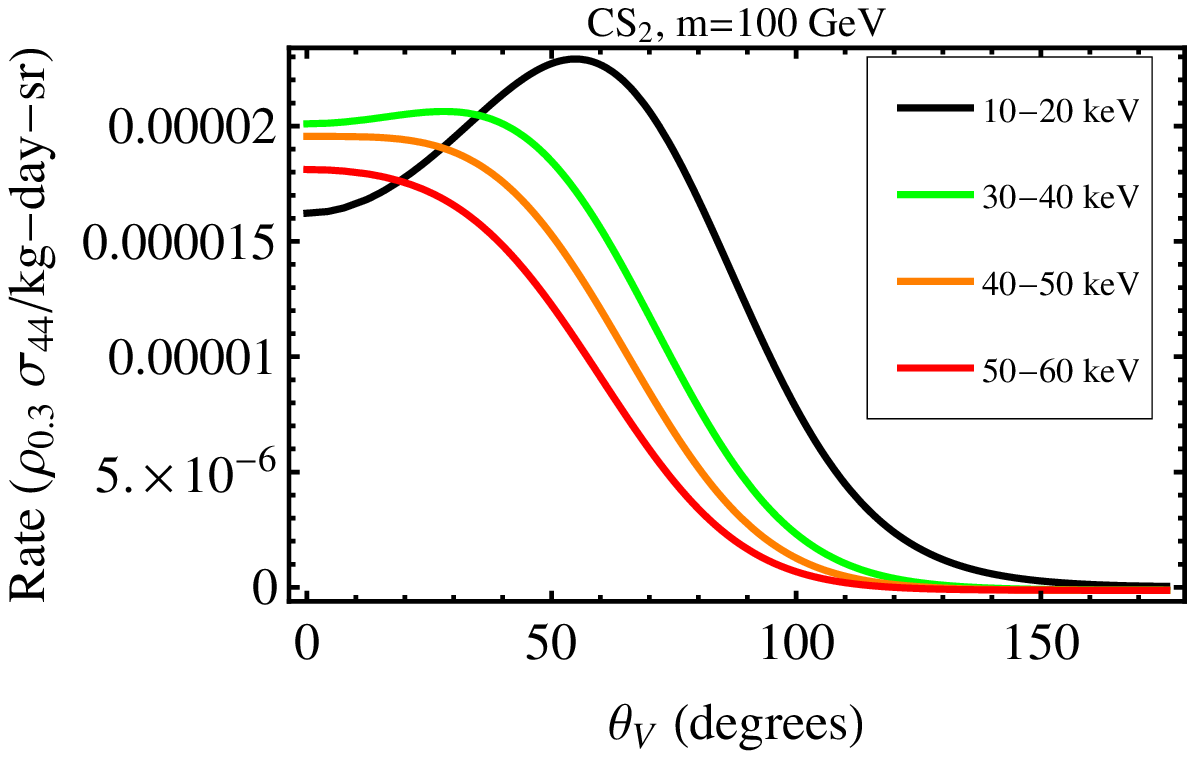}
  \includegraphics[height=130pt]{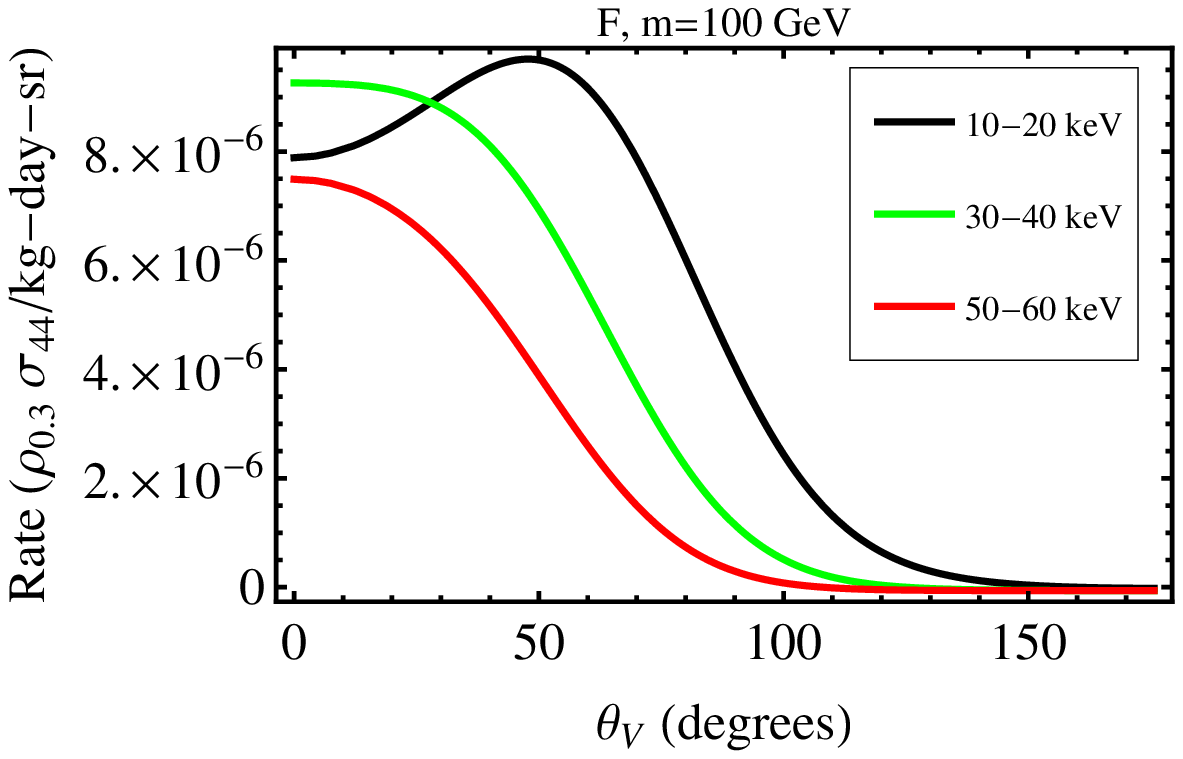}\\
  \vspace{-0.1cm}\caption{Same as Fig.~\ref{RateAngle-CS2-Integrated}.a but for 10 keV energy intervals in (a) CS$_2$ and (b) F. The black, green, orange (only in (a)) and red curves are for $10~{\rm keV} \leq E_R \leq 20$ keV,  $30~{\rm keV} \leq E_R \leq 40$ keV, $40~{\rm keV} \leq E_R \leq 50$ keV and  $50~{\rm keV} \leq E_R \leq 60$ keV, respectively.}
  \label{RateAngle-Integrated-DiffBin}
\end{center}
\end{figure}

Fig.~\ref{RateAngle-CS2-Integrated}.a shows the CS$_2$ energy-integrated directional rate,
\begin{equation}
\mathcal{R}(E_1, E_2; \theta_V)=\int_{E_1}^{E_2}{\frac{dR}{dE_R~d\phi_V~d\cos\theta_V}\Big|_{\phi_V} dE_R}=\frac{dR(E_1,E_2)}{d\phi_V~d\cos\theta_V}\Big|_{\phi_V},
\label{EnergyIntRate}
\end{equation}
at fixed azimuthal angle $\phi_V$ integrated over different energy intervals as a function of $\theta_V$ for $m=100$ GeV/$c^2$, $V_{\rm lab}=312$ km/s and $\sigma_v=173$ km/s on June 2. In Fig.~\ref{RateAngle-CS2-Integrated}.b, the integrated recoil rate has been re-scaled to be between 0 and 1, what makes it easier to see that rings exist in the rate integrated from 5 keV up to 60 keV and lower energies.
Fig.~\ref{RateAngle-Integrated-DiffBin} shows $\mathcal{R}$ for  different 10 keV energy intervals instead, always as a function of $\theta_V$ and for the same parameters of Fig.~\ref{RateAngle-CS2-Integrated},  again for CS$_2$ (left panel),  and also for F (right panel). Fig.~\ref{RateAngle-Integrated-DiffBin}.b is similar to Fig.~2 of Ref.~\cite{Billard:2009mf} (in which no ring feature is seen), but for different values of $V_{\rm lab}$ and $\sigma_v$.

From Figs.~\ref{RateAngle-CS2-Integrated} and \ref{RateAngle-Integrated-DiffBin} we can conclude that a ring can be observed in the integrated rate, but the energy threshold should be as low as possible.

\section{Statistical significance of the ring}

Here we use a simple statistical test to estimate the number of events needed to detect the ring assuming no background (and not taking into account the energy and angular experimental resolutions). We define two angular regions centered at $-{\bf V}_{\rm lab}$: the ``ring'' and the ``center''. The ``ring'' is defined as the region with $\gamma_1 < \theta_V < \gamma_2$. The ``center'' is defined as the $\theta_V<\gamma_1$ region. The angles $\gamma_1$ and $\gamma_2$, which determine the thickness of the ring, are in principle to be chosen in an optimal way. Here, for simplicity, we take them to be the $\theta_V$ values for which the energy-integrated directional rate (at fixed azimuthal angle $\phi_V$), $\mathcal{R}(E_1, E_2; \theta_V)$ (Eq.~\ref{EnergyIntRate}) is halfway between its value at $-{\bf V}_{\rm lab}$ (i.e.~ $\theta_V=0$) and its maximum value (at $\theta_V=\gamma$),
\begin{equation}
\mathcal{R}(E_1,E_2;\gamma_1)=\mathcal{R}(E_1,E_2;\gamma_2)=\frac{\mathcal{R}(E_1,E_2;0)+\mathcal{R}(E_1,E_2;\gamma)}{2}.
\label{RingThickness}
\end{equation}
The solid angle subtended by the ring and the center regions so defined are $\Omega_r=2 \pi (\cos \gamma_1 - \cos \gamma_2)$ and  $\Omega_c=2 \pi (1 - \cos \gamma_1)$.

To observe the ring at the 3$\sigma$ level, the difference between the number of events in the ring and the center regions, $N_r$ and $N_c$, should be large enough. At the 3$\sigma$ level, then
\begin{equation}
N_r-N_c > 3\sigma \simeq 3 \sqrt{N},
\label{RingCond}
\end{equation}
where $N = N_r + N_c$ is the total number of events for $\theta_V<\gamma_2$ over the duration of the experiment. The number of events needed depends on the energy interval. Assuming a particular interval $(E_1,E_2)$, we approximate $N_r$ and $N_c$ using the middle values of $\mathcal{R}$ (Eq.~\ref{EnergyIntRate}) in each region,
\begin{equation}
N_r(E_1, E_2)=\frac{\mathcal{R}(\gamma_1)+\mathcal{R}(\gamma_2)}{2}~\Omega_r~MT \equiv \overline{\mathcal{R}}_r(E_1,E_2)~\Omega_r~MT,
\label{Nr}
\end{equation}
\begin{equation}
N_c(E_1, E_2)=\frac{\mathcal{R}(0)+\mathcal{R}(\gamma_1)}{2}~\Omega_c~MT \equiv \overline{\mathcal{R}}_c(E_1,E_2)~\Omega_c~MT.
\label{Nc}
\end{equation}
$M T$ is the detector exposure. To exemplify this procedure, Fig.~\ref{RingThickness} shows the ring thickness limited by $\gamma_1$ and $\gamma_2$, and the rates $\overline{\mathcal{R}}_{\rm ring}$ and $\overline{\mathcal{R}}_{\rm center}$ with dashed horizontal lines in the integrated directional recoil rate in the 5-40 keV interval reproduced from Fig.~\ref{RateAngle-CS2-Integrated}.a.
  \begin{figure}[t]
\begin{center}
  \includegraphics[height=200pt]{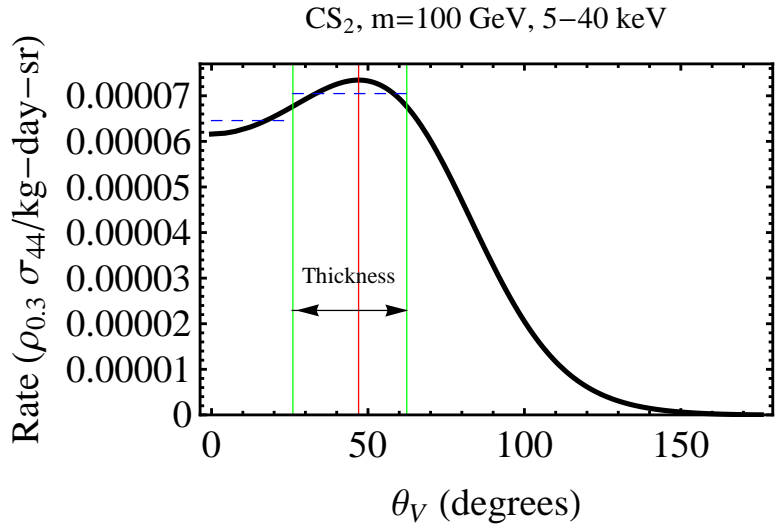}
  \vspace{-0.1cm}\caption{Ring radius at $\theta_V=\gamma$ (central red vertical line) and  thickness of the ``ring" region (between the two lateral green vertical lines), which in this case extends from $\gamma_1=26^\circ$ to $\gamma_2=62^\circ$,  for the energy integrated directional rate in the 5 - 40 keV interval reproduced from  Fig.~\ref{RateAngle-CS2-Integrated}.a. The dashed blue horizontal lines in the ``central" and ``ring" regions specify $\overline{\mathcal{R}}_{\rm ring}$, and $\overline{\mathcal{R}}_{\rm center}$ (Eqs.~\ref{Nr} and \ref{Nc}), respectively.}
  \label{RingThickness}
\end{center}
\end{figure}

From Eq.~\ref{RingCond} to \ref{Nc} we can eliminate the exposure $MT$ (which depends on $\sigma_p$ and $\rho$) and find $N$ for the particular $(E_1,E_2)$ energy range (for $\theta_V<\gamma_2$),
\begin{equation}
N(E_1,E_2) > \left(3~ \frac{\overline{\mathcal{R}}_r(E_1,E_2)~\Omega_r + \overline{\mathcal{R}}_c(E_1,E_2)~\Omega_c}{\overline{\mathcal{R}}_r(E_1,E_2)~\Omega_r - \overline{\mathcal{R}}_c(E_1,E_2)~\Omega_c} \right)^2 =N_{\rm min}(E_1,E_2).
\label{RingCond-N}
\end{equation}
Notice that the larger the difference in the solid angles $\Omega_r$ and $\Omega_c$, the smaller is the needed $N_{\rm min}$.
Using the minimum number of events $N_{\rm min}(E_1,E_2)$ and Eqs.~\ref{Nr} and \ref{Nc}, we can find the minimum necessary exposure
$(MT)_{\rm min}$ in units of kg-yr/$\rho_{0.3} \sigma_{44}$, which allows us to find the minimum number of events in all directions $N_{\rm tot}$ (in any energy interval $(E'_1, E'_2)$),
\begin{equation}
N_{\rm tot}(E'_1, E'_2)=(MT)_{\rm min}~2\pi~\int{\mathcal{R}(E'_1, E'_2;\theta_V) d\cos\theta_V} \equiv (MT)_{\rm min} \mathcal{R}_{\rm tot}(E'_1, E'_2).
\label{RingCond-Ntot}
\end{equation}
Notice that $N_{\rm tot}$ is independent of $\sigma_p \rho$, as it can also be seen from an alternative expression for it, which avoids using the exposure,
\begin{equation}
N_{\rm tot}(E'_1, E'_2)=N\left(\frac{\overline{\mathcal{R}}_r(E_1,E_2)~\Omega_r}{\mathcal{R}_{\rm tot}(E'_1, E'_2)} + \frac{\overline{\mathcal{R}}_c(E_1,E_2)~\Omega_c}{\mathcal{R}_{\rm tot}(E'_1, E'_2)}\right)^{-1}.
\label{RingCond-Ntot-2}
\end{equation}

For CS$_2$ and $m=100$ GeV/$c^2$, $V_{\rm lab}=312$ km/s and $\sigma_v=173$ km/s, in the 5-40 keV interval (see curve in Fig.~\ref{RateAngle-CS2-Integrated}), we find $N_{\rm min}=21$ events for $\theta_V<62^{\circ}$, i.e.~$N_{\rm tot}=41$ events over all directions ($\overline{\mathcal{R}}_{\rm center}/\overline{\mathcal{R}}_{\rm ring}=0.92$, $MT=250$ kg-yr/$\rho_{0.3} \sigma_{44}$) are needed to detect the ring. In the 20-40 keV interval we would instead need $N_{\rm tot}=55$ events ($\overline{\mathcal{R}}_{\rm center}/\overline{\mathcal{R}}_{\rm ring}=0.97$, $MT=700$ kg-yr/$\rho_{0.3} \sigma_{44}$). For this WIMP mass, the ring disappears when integrating over energies above 40 keV. With this last exposure, there would be 136 events above 20 keV (but the ring could not be seen if integrating above 40 keV). This is about 5 times more than needed to detect the mean recoil direction, according to Ref.~\cite{Green:2010gw}.

With our simple method we can also estimate the number of events required to detect the forward-backward asymmetry $N_{\rm FB}$, using an equation similar to Eq.~\ref{RingCond-N}. In this case, $\gamma_1=\pi/2$, and $\gamma_2=\pi$, thus $\Omega_{\rm forward}=\Omega_{\rm backward}=2\pi$. $\overline{\mathcal{R}}_{\rm forward}$ and $\overline{\mathcal{R}}_{\rm backward}$ are defined here as
\begin{eqnarray}
\overline{\mathcal{R}}_{\rm forward}(E_1,E_2)&=&\frac{\mathcal{R}(0)+\mathcal{R}(\pi/2)}{2},\nonumber\\
\overline{\mathcal{R}}_{\rm backward}(E_1,E_2)&=&\frac{\mathcal{R}(\pi/2)+\mathcal{R}(\pi)}{2},
\label{Rbar-FB}
\end{eqnarray}
where the rates $\mathcal{R}$ are also defined between $E_1$ and $E_2$ (Eq.~\ref{EnergyIntRate}). Thus
\begin{equation}
N_{\rm FB} > \left(3~ \frac{\overline{\mathcal{R}}_{\rm forward} + \overline{\mathcal{R}}_{\rm backward}}{\overline{\mathcal{R}}_{\rm forward} - \overline{\mathcal{R}}_{\rm backward}} \right)^2 .
\label{RingCond-NFB}
\end{equation}

For CS$_2$ and  $m=100$ GeV/$c^2$, we find that $N_{\rm FB}=17$ events would be required above 5 keV to detect the forward-backward asymmetry ($\overline{\mathcal{R}}_{\rm backward}/\overline{\mathcal{R}}_{\rm forward}=0.16$, and the exposure needed would be 57 kg-yr/$\rho_{0.3} \sigma_{44}$). For $E_R>20$ keV instead, we find $N_{\rm FB}=13$ events ($\overline{\mathcal{R}}_{\rm backward}/\overline{\mathcal{R}}_{\rm forward}=0.09$, $MT=56$ kg-yr/$\rho_{0.3} \sigma_{44}$). This number, obtained with our simple method, compares well with the number of events required to reject isotropy, 9 events, in Ref.~\cite{Green:2010gw}.

\section{WIMP mass determination}

Several ways of extracting the WIMP mass from direct detection data have been explored. The WIMP mass could be extracted from the energy spectrum~\cite{Green:2008, Shan:2009, Schnee:2011} and from the mean recoil energy~\cite{Schnee:2011} in non-directional detection,  from the directional rate as a function of $\theta_V$~\cite{Billard:2009mf}, or from both the angular and energy distributions~\cite{Billard:2011, Lee:2012} in directional direct detection. The ring angular radius $\gamma$ in the differential rate (not integrated over energy) may lead to an additional indication of the WIMP mass, since it does not exist for very small masses.

For a given recoil energy and target element, one could obtain some information on the mass of the WIMP from the angular radius of the ring. If a ring is present in the differential recoil rate at the energy $E_R$, for a target nucleus of mass $M$ one can find $m=\mu M/(M-\mu)$ where
\begin{equation}
\mu=\frac{1}{\cos \gamma} \sqrt{\frac{ME_R}{2 V_{\rm lab}^2}}.
\label{mu-ring}
\end{equation}
The $m$ so derived should be the same for all $E_R$ (i.e. $\cos \gamma \propto \sqrt{E_R}$). If a ring does not exist, the maximum recoil rate is in the direction of  $-{\bf V}_{\rm lab}$. The ring exists only if the reduced mass $\mu$ is in the range
\begin{equation}
\sqrt{\frac{ME_R}{2 V_{\rm lab}^2}} \leq \mu \leq M.
\label{mu-ring-range}
\end{equation}

\begin{figure}[t]
\begin{center}
  \includegraphics[height=140pt]{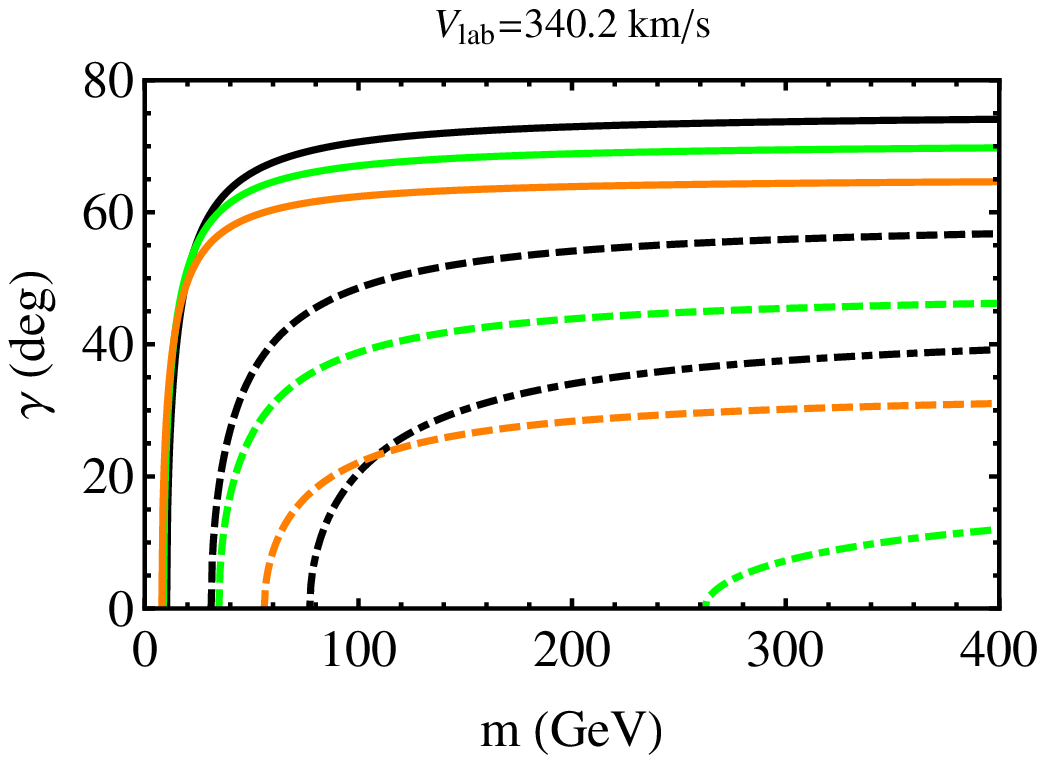}
  \includegraphics[height=140pt]{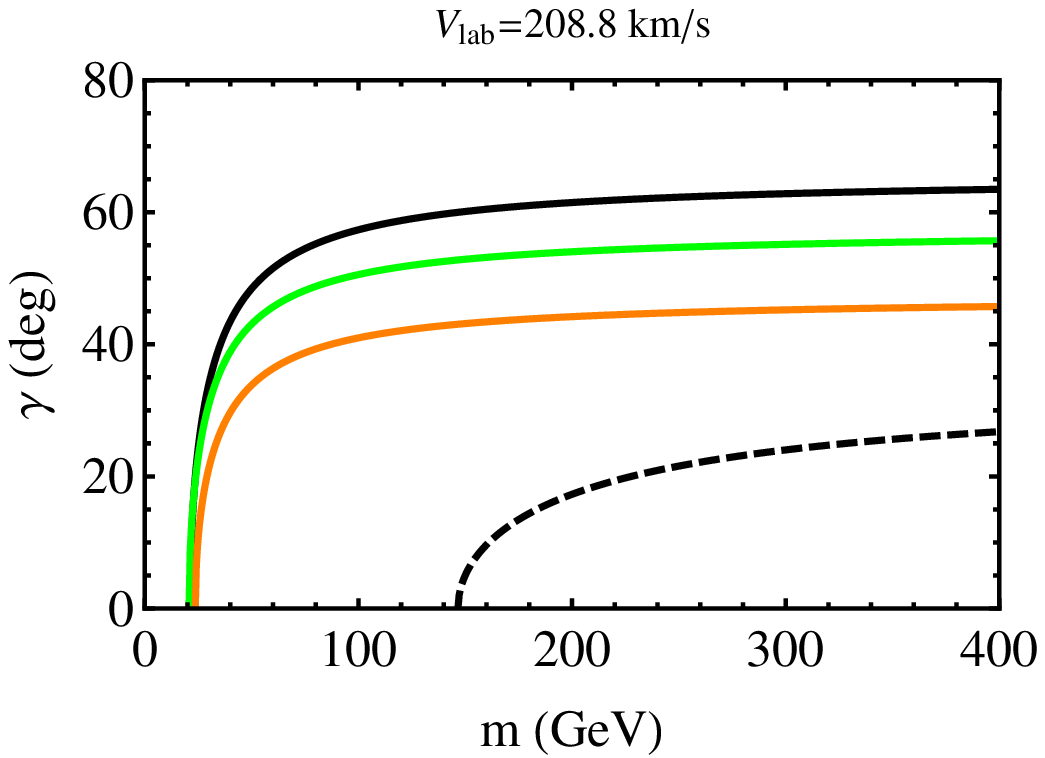}\\
  \vspace{-0.1cm}\caption{Ring radius $\gamma$, measured in degrees from the direction of $-{\bf V}_{\rm lab}$, as a function of $m$ for three target elements S (black lines), F (green lines), and C (orange lines) and three recoil energies $E_R=5$ keV (solid lines), 20 keV (dashed lines), and 40 keV (dot-dashed lines) for (a) $V_{\rm lab}=340.2$ km/s, and (b) $V_{\rm lab}=208.8$ km/s. Above a certain value of $m$, $\gamma$ reaches a plateau.}
  \label{gammaVsmass}
\end{center}
\end{figure}

Fig.~\ref{gammaVsmass} shows a plot of $\gamma$ in the differential rate as  a function of $m$ (from Eq.~\ref{mu-ring}) for S, F, and C at three different recoil energies of 5, 20, and 40 keV. The two panels of Fig.~\ref{gammaVsmass} illustrate the impact of different values of $V_{\rm GalRot}$, 312 km/s and 180 km/s, so that  $V_{\rm lab}=340.2$ km/s and 208.8 km/s, respectively. Due to Eq.~\ref{maxgamma}, if $m \gg M$ there is no sensitivity of $\gamma$ to the WIMP mass $m$, and $\gamma$ reaches a plateau (see Fig.~\ref{gammaVsmass}).

\begin{figure}[t]
\begin{center}
  \includegraphics[height=200pt]{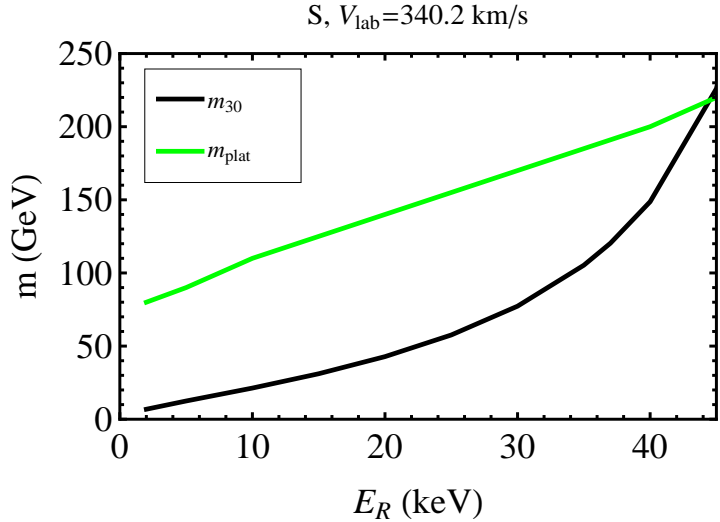}
  \vspace{-0.1cm}\caption{Sensitivity region where a ring radius can be used to indicated a WIMP mass. The range of $m$ is shown between $m_{30}$ (black curve), below which $\gamma \leq 30^\circ$, and $m_{\rm plat}$ (green curve) above which $\gamma$ reaches a plateau (thus $m \geq m_{\rm plat}$), as a function of $E_R$ for $V_{\rm GalRot}=312$ km/s in an S detector.}
 \label{MassRange}
\end{center}
\end{figure}

It would be difficult to observe a ring with angular radius in the directional rate $\gamma < 30^\circ$ (the rates at the center and  the ring are almost equal), so only for the range of $m$ shown in Fig.~\ref{MassRange} as a function of $E_R$ (between $m_{30}$, below which $\gamma \leq 30^\circ$, and $m_{\rm plat}$ above which $\gamma$ reaches a plateau), a measurement of $\gamma$ could translate into a measurement of $m$. One cannot distinguish $m_{\rm plat}$ from any mass larger than it, $m \geq m_{\rm plat}$, using the ring radius. Fig.~\ref{MassRange} corresponds to an S detector and  $V_{\rm lab}=340.2$ km/s. Similar ranges can be easily found for F and C using Fig.~\ref{gammaVsmass}.

Eq.~\ref{RingCond-N} can also be used to find the number of events needed to observe the ring in the directional differential rate. If we take a narrow enough energy bin of width $\Delta E_R$  in keV so that the differential rate would be constant in the bin, Eq.~\ref{EnergyIntRate} becomes
\begin{equation}
\mathcal{R}(E, E+\Delta E_R;\theta_V)=\frac{dR}{dE_R~d\phi_V~d\cos\theta_V}\Big|_{\phi_V} \Delta E_R.
\label{EnergyIntRate-DeltaE}
\end{equation}
We choose here a bin $\Delta E_R \simeq 1$ keV. For CS$_2$, $m=100$ GeV/$c^2$, $E_R=5$ keV, $V_{\rm lab}=312$ km/s, and $\sigma_v=173$ km/s (shown in Fig.~\ref{RateAngle-CS2}.a), the exposure needed to observe the ring in the differential rate is $4.0 \times 10^3$ kg-yr-keV/$\rho_{0.3} \sigma_{44} \Delta E_R$. This corresponds to 19 events needed in a 1 keV energy bin at 5 keV within 95$^{\circ}$ of $-{\bf V}_{\rm lab}$ to observe the ring, namely 650 total events (over all directions) in the 5-40 keV interval, or 780 events above 20 keV. For the same parameters but at $E_R=20$ keV, a minimum of 20 events would be required within 64$^{\circ}$ of $-{\bf V}_{\rm lab}$ to observe the ring which corresponds to an exposure and total number of events higher by about a factor of two than those just mentioned.

We conclude that detecting the ring in the differential rate would require number of events at least 6 times larger that those required using the energy integrated rate, i.e. 30  times more than needed to detect the mean  recoil direction according to Ref.~\cite{Green:2010gw}.

\section{Anisotropic logarithmic-ellipsoidal models}

So far we used the isotropic Maxwell-Boltzmann model but here we will show that  the ring-like feature persists with different characteristics in an anisotropic halo. Various observations and numerical simulations suggest that triaxial models with anisotropic velocity distributions are a better approximation for galaxy halos~\cite{Moore:2001, Helmi:2002, Green:2002}. In the anisotropic logarithmic-ellipsoidal model of Ref.~\cite{Evans:2000}, the recoil momentum distribution is anisotropic  and can be written as~\cite{Alenazi-Gondolo:2008}
\begin{equation}
\hat{f}_{\rm lab}\!\left(\frac{q}{2\mu}, \hat{\bf q} \right)=\frac{1}{\sqrt{2\pi (\sigma_{xg}^2 q_{xg}^2 + \sigma_{yg}^2 q_{yg}^2 + \sigma_{zg}^2 q_{zg}^2)}} \times \exp{\left(-\frac{\left[(q/2\mu) + \hat{\bf q} . {\bf V}_{\rm lab}\right]^2}{2(\sigma_{xg}^2 q_{xg}^2 + \sigma_{yg}^2 q_{yg}^2 + \sigma_{zg}^2 q_{zg}^2)}\right)},
\label{fhat-anisotropic}
\end{equation}
where $q_{xg}$, $q_{yg}$, and $q_{zg}$ are components of $\hat{\bf q}$ in galactic coordinates (see Appendix A), and the velocity dispersion matrix in the galactic reference frame is given by,
\begin{equation}
\boldsymbol\sigma_v^2=
\left( \begin{array}{ccc}
\sigma_{xg}^2 & 0 & 0 \\
0 & \sigma_{yg}^2 & 0 \\
0 & 0 & \sigma_{zg}^2 \end{array} \right).
\end{equation}
Since we assume that the WIMPs are on average at rest with respect to the Galaxy, the average WIMP velocity $- {\bf  V}_{\rm lab}$ in  this model is the same as for the IMB. The equations for $\sigma_{xg}$, $\sigma_{yg}$, and $\sigma_{zg}$, as well as the constants $\mathfrak{p}$ and $\mathfrak{q}$ used to describe the axis ratios of the density ellipsoid, and the constant $\Gamma$ used to describe the velocity anisotropy, are given in Appendix B.

\begin{figure}[t]
\begin{center}
  \includegraphics[height=90pt]{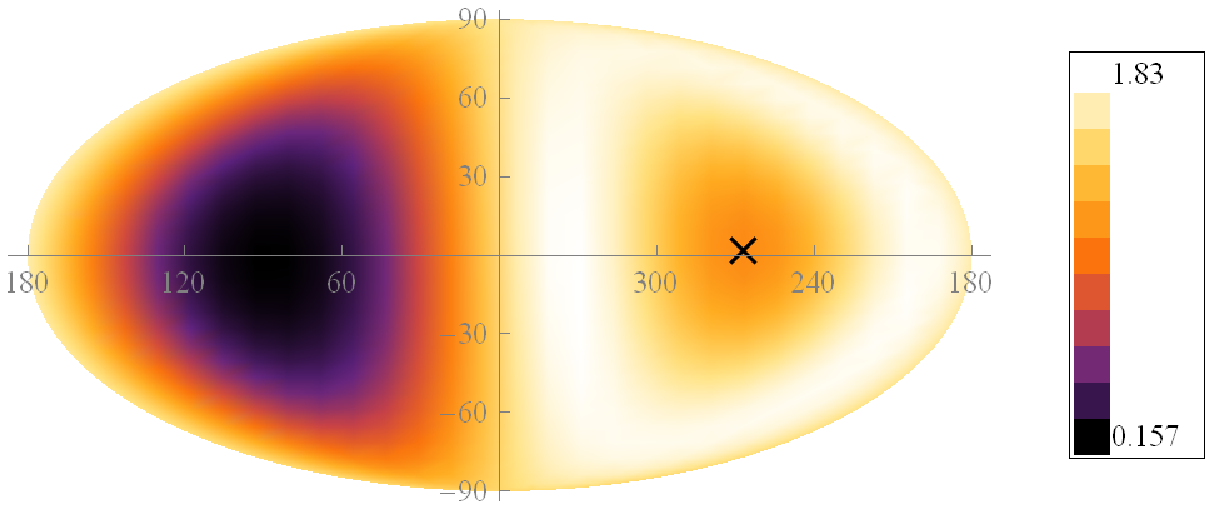}
  \includegraphics[height=90pt]{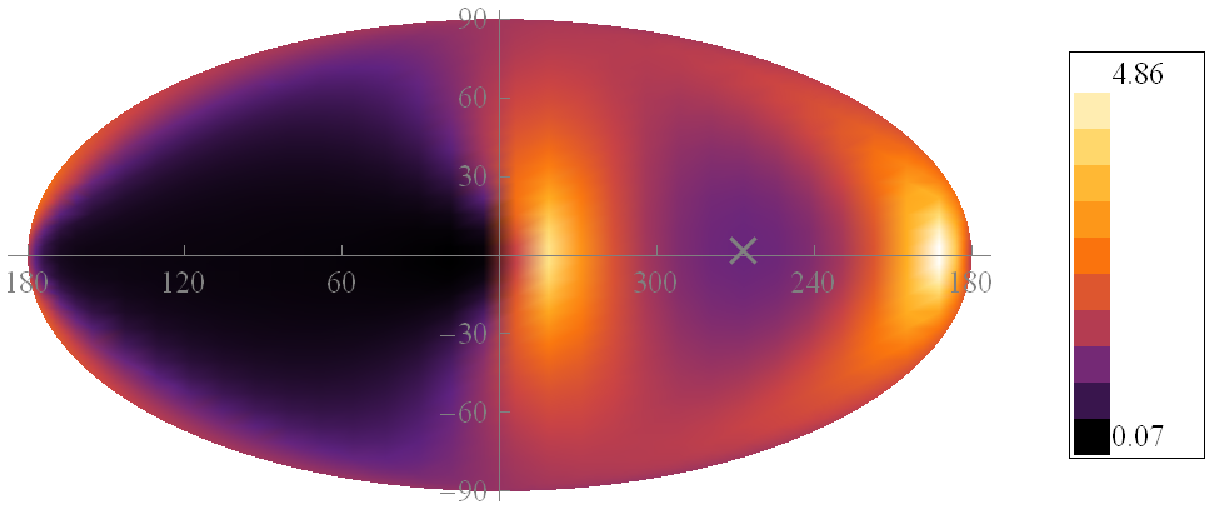}
  \includegraphics[height=90pt]{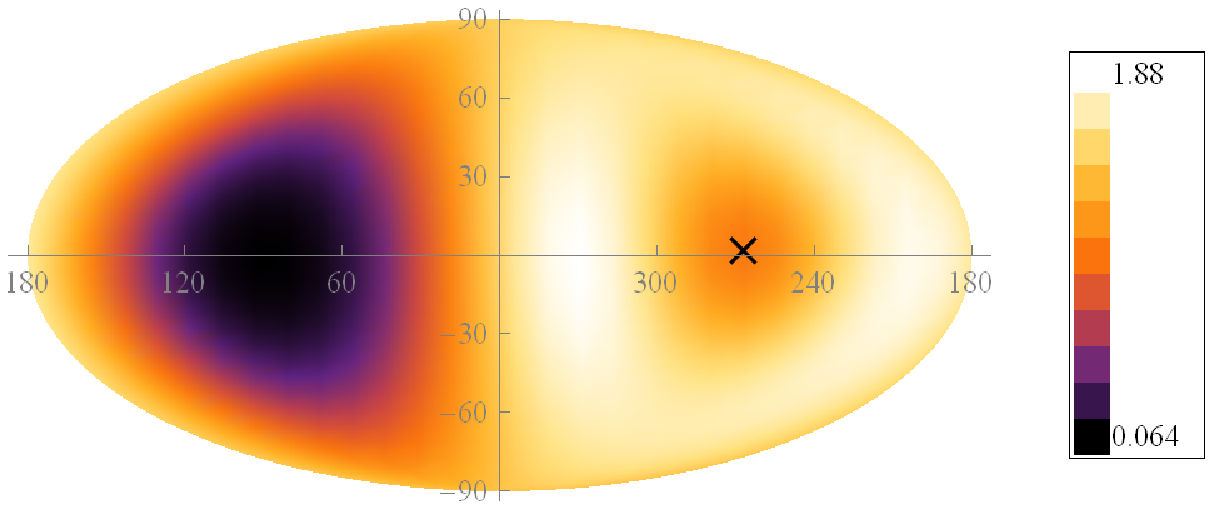}
  \includegraphics[height=90pt]{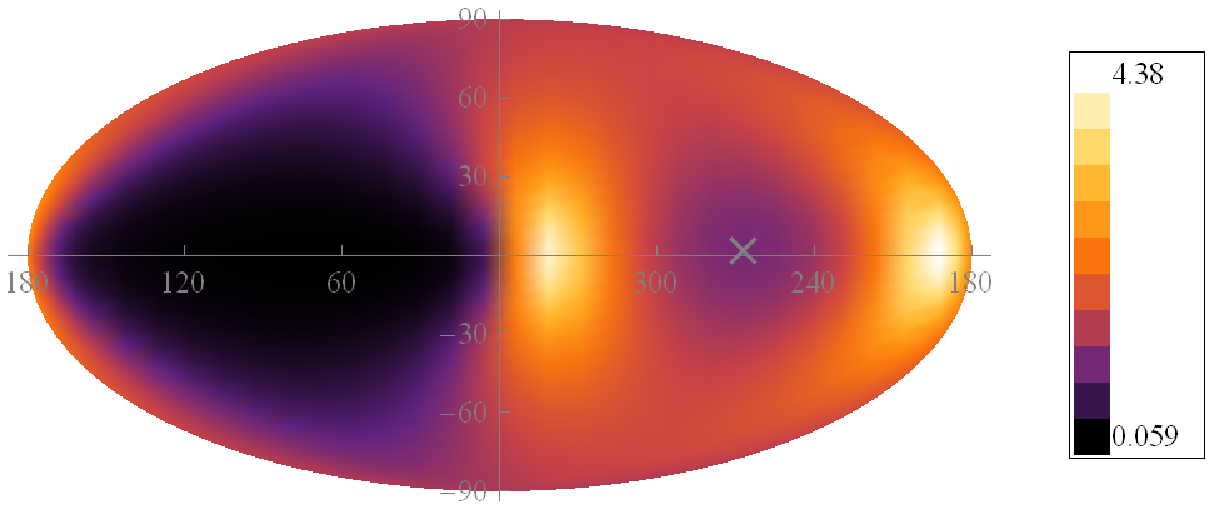}\\
  \vspace{-0.1cm}\caption{Directional differential recoil rate for $m = 100$ GeV/$c^2$, $E_R=5$ keV in CS$_2$, for the anisotropic logarithmic-ellipsoidal model with $\mathfrak{p}=0.9$, $\mathfrak{q}=0.8$,  $v_{\rm esc}=544$ km/s and $V_{\rm GalRot}=312$ km/s on June 2. Top left (a): Solar system on the major axis and $\Gamma=-1.0$. Top right (b): Solar system on the major axis and $\Gamma=16$. Bottom left (c): Solar system on the minor axis and $\Gamma=0.07$. Bottom right (d):Solar system on the minor axis and $\Gamma=16$. The color scale/grayscale shown in the vertical bars correspond to equal steps between the minimum and maximum values given in units of $10^{-6} \times (\rho_{0.3} \sigma_{44}/{\text{kg-day-keV-sr}})$. The ring is clearly visible and similar to the case of the IMB in (a) and (c). In (b) and (d), $|\Gamma|$ is large and the rate is different at different azimuthal angles around $-{\bf V}_{\rm lab}$ and is harder to see the ring.}
  \label{Mollweide-Anisot}
\end{center}
\end{figure}

Eq.~\ref{fhat-anisotropic} shows that the angular radius of the ring in this anisotropic model is identical to that of the IMB in Eq.~\ref{gamma}. The ring is still circular with constant radius at any azimuthal angle $\phi_V$ around $-{\bf V}_{\rm lab}$, but the value of the rate at the maximum ($\theta_V=\gamma$) depends on $\phi_V$. In the left Mollweide maps of Fig.~\ref{Mollweide-Anisot} the ring is almost indistinguishable from the ring in the IMB, but in the right plots of the same figure there is a strong azimuthal variation.

Keeping $\mathfrak{p}$ and $\mathfrak{q}$ constant, the change of the rate at different azimuthal angles is larger for larger values of $|\Gamma|$. This can be seen in Fig.~\ref{Mollweide-Anisot} where we show a Mollweide map of the directional differential recoil rate for $m = 100$ GeV/$c^2$, $E_R=5$ keV in CS$_2$, using the Maxwellian distribution in Eq.~\ref{fhat-anisotropic} with $v_{\rm esc}=544$ km/s and $V_{\rm GalRot}=312$ km/s (on June 2). The Solar system is on the major and minor axes in the top and bottom plots, respectively. For all four plots, $\mathfrak{p}=0.9$ and $\mathfrak{q}=0.8$. In the top row, $\Gamma=-1.0$ and 16 for the left and right plots, respectively, and in the bottom row $\Gamma=0.07$ and 16 for the left and right plots, respectively. The left plots with smaller $|\Gamma|$ are similar to those in the IMB with similar $V_{\rm lab}$ (Figs.~\ref{Flux}.b and \ref{VGalRot-sigmav}.c). The right plots with large $|\Gamma|$ are very anisotropic with respect to $-{\bf V}_{\rm lab}$. Thus, the ring-like feature can give information on the WIMP velocity distribution. More work would be necessary to clarify this issue.

\section{Discussion}

We have pointed out that at low enough recoil energies and for heavy enough WIMPs, the maximum of the recoil  rate is not in the direction opposite to the velocity of the detector with respect to the galaxy, $-{\bf V}_{\rm lab}$, but on a ring around it at an angle $\gamma$ that approaches 90$^\circ$ at very low energies.

For the nuclei of interest in direct detection experiments (at present S, C and F), the maximum possible values of $\gamma$ are shown in Fig.~\ref{cosgamma}. The  ring radius $\gamma$ is larger for larger values of the average velocity of the WIMPs with respect to the detector.

The ring-like feature can be used as a secondary signature of dark matter since the background rate would not have such a feature. The minimum number of events required to detect the ring in the energy-integrated rate over relatively low recoil energies is larger than required to detect the mean recoil direction, but possible by a factor as low as five, if low threshold energies can be achieved in directional detectors.

In Section 3, for one of the most favorable sets of  halo parameters we studied, for  a 100 GeV/$c^2$ WIMP we found with a simple statistical method that 41 events with energies between 5 keV and 40 keV, or 55 events between 20 keV and 40 keV, would be required to detect the ring in the energy-integrated directional recoil rate in CS$_2$.  With the corresponding exposure of 700 kg-yr/$\rho_{0.3} \sigma_{44}$, there would be 136 events  in CS$_2$ above 20 keV (but the ring could not be seen if integrating above 40 keV  for  a 100 GeV/$c^2$ mass WIMP).

This number of events necessary to observe the ring, 136, is about 5 times larger than  the approximately 30 events needed to determine the mean recoil direction according  to Ref.~\cite{Green:2010gw} in the same energy range, with the same WIMP mass and detector material. Using the same simple statistical method, we estimated that 13 events would be required to detect the forward-backward asymmetry for recoil energies above 20 keV in the same case. This result compares well with the approximately 10 events in the same energy range that were found in  Refs.~\cite{Morgan:2005} and~\cite{Green:2010gw} to be necessary (using  a much more sophisticated statistical method).

As explained in Section 4, some information can be obtained on the WIMP mass from the existence and the angular radius of the ring in the directional differential recoil rate, but this would require a number of events at least 6 times larger than those necessary  when using the energy-integrated directional rate. For example for the most favorable halo parameters and a 100 GeV/$c^2$ WIMP, with the exposure required to observe the ring in  a 1 keV energy bin at 5 keV  in CS$_2$, 650 events in the 5-40 keV interval, or 780 events above 20 keV would be needed. To study the expected precision of the mass determination requires Monte-Carlo studies outside the scope of this paper.

The present threshold of DRIFT  is 40 keV. With this threshold, the ring is present in S recoils only for WIMPs with $m \geq 600$ GeV/$c^2$, for which  the recoil rates are smaller and large number of events  would be needed to observe the ring. For a 600 GeV/$c^2$ WIMP and assuming the most favorable halo parameters we studied, 53 events between 40 keV and 50 keV, which means an exposure of $8.9 \times 10^3$ kg-yr/$\rho_{0.3} \sigma_{44}$, would be required to detect the ring in  CS$_2$. With this exposure, there would be 315 events above 40 keV.

We have assumed all along that the recoil directions, including their senses, can be reconstructed perfectly in 3 dimensions and there are no backgrounds. Assuming more realistic conditions will make the ring-like feature more difficult to observe.

In conclusion, the ring is the place where most of the lower energy events would come from in directional detectors, if the WIMP mass is large enough. This feature can be used as a secondary signature for dark matter, given that it requires more events to be detected than  other signatures already discussed in the literature. The detailed features of the ring depend on the WIMP mass and velocity distribution. Further study is needed to find the optimal strategy and the best achievable precision in the determination of the WIMP mass using the ring.
 The ring-like feature could also be used to test the shape of the WIMP velocity distribution. For example, we showed that logarithmic-ellipsoidal distributions give an azimuthal pattern around the ring, while the ring remains circular. Further study is required to clarify this issue too.

\begin{acknowledgments}
G.G. and N.B.  were supported in part by the US Department of Energy Grant
DE-FG03-91ER40662, Task C.  P.G. was  supported  in part by  the NFS
grant PHY-1068111 at the University of Utah. P.G. thanks the Oskar Klein Centre at the University of Stockholm for support during his sabbatical visit.
\end{acknowledgments}

\appendix
\addappheadtotoc

\section{Complete transformation equations from the detector frame to the Galactic reference frame}

The presence of $\hat{\mathbf{q}} \cdot {\bf V}_{\rm {lab}}$ in $\hat{f}_{\rm lab}$ means that in order to compute the differential rate we need to orient the nuclear recoil direction $\hat{\mathbf{q}}$ with respect to ${\bf V}_{\rm {lab}}$. We present here complete transformation equations for $\hat{\mathbf{q}}$ and ${\bf V}_{\rm {lab}}$ to go from the detector frame to the Galactic frame.

We define a reference frame fixed to the laboratory and orient its axes so that the $xy$ plane is horizontal, the $x$-axis points North, the $y$-axis points West, and the $z$-axis points to the Zenith. We denote its unit coordinate vectors as $\hat{\cal N}$, $\hat{\cal W}$ and $\hat{\cal Z}$, respectively. The detector is at some orientation in the laboratory. We define the detector frame with $X,Y,Z$ cartesian axes fixed to the detector. The unit coordinate vectors of the detector frame are $\hat{\mathbf{X}}$, $\hat{\mathbf{Y}}$ and $\hat{\mathbf{Z}}$. The transformation between the lab frame and the detector frame is given in Eqs. A1 to A3 of Ref.~\cite{ChanIV} in terms of ``direction cosines''. $\hat{\mathbf{q}}$ is given in the detector reference frame as $\hat{\mathbf{q}}=q_X ~\hat{\mathbf{X}} +q_Y ~\hat{\mathbf{Y}} +q_Z ~\hat{\mathbf{Z}}$, and we can write it in the lab frame using Eq. A1 of Ref.~\cite{ChanIV}, $\hat{\mathbf{q}}=q_n ~\hat{\cal N}+q_w ~\hat{\cal W} +q_z ~\hat{\cal Z}$, where
\begin{align}
q_n&=q_X \alpha_X + q_Y \alpha_Y +q_Z \alpha_Z,\nonumber\\
q_w&=q_X \beta_X + q_Y \beta_Y +q_Z \beta_Z,\nonumber\\
q_z&=q_X \gamma_X + q_Y \gamma_Y +q_Z \gamma_Z,
\end{align}
and $\alpha_i$, $\beta_i$ and $\gamma_i$ are the direction cosines between the two sets of cartesian coordinates of the lab and detector frames, for $i=X,Y,Z$. For example $\alpha_Y$ is the cosine of the angle between the $\hat{\cal N}$ and $\hat{\mathbf{Y}}$ directions, and $\beta_Z$ is the cosine of the angle between the $\hat{\cal W}$ and $\hat{\mathbf{Z}}$ directions.

We would like to write $\hat{\mathbf{q}}$ in the Galactic reference frame. Using Eq. A8 of Ref.~\cite{ChanIV}, $\hat{\mathbf{q}}$ can be written in the equatorial frame,
\begin{align}
\hat{\mathbf{q}} &= \left[-q_n \sin(\lambda_{\rm lab})\cos (t^\circ_{\rm lab}) +q_w \sin (t^\circ_{\rm lab}) + q_z \cos(\lambda_{\rm lab})\cos (t^\circ_{\rm lab}) \right] \hat{\bf x}_e \nonumber\\
&+\left[-q_n \sin(\lambda_{\rm lab})\sin (t^\circ_{\rm lab}) -q_w \cos (t^\circ_{\rm lab}) + q_z \cos(\lambda_{\rm lab})\sin (t^\circ_{\rm lab}) \right] \hat{\bf y}_e \nonumber\\
&+\left[q_n \cos(\lambda_{\rm lab}) + q_z \sin(\lambda_{\rm lab}) \right] \hat{\bf z}_e,
\label{q-Equit}
\end{align}
where $\hat{\bf x}_e$,  $\hat{\bf y}_e$, and  $\hat{\bf z}_e$ are the unit coordinate vectors of the geocentric equatorial inertial (GEI) frame: its origin is at the center of the Earth, its $x_e$-axis points in the direction of the vernal equinox, its $y_e$-axis points to the point on the celestial equator with right ascension 90$^\circ$ (so that the cartesian frame is right-handed), and its $z_e$-axis points to the north celestial pole. $t^\circ_{\rm lab}$ is the laboratory Local Apparent Sidereal Time in degrees, and $\lambda_{\rm lab}$ is the latitude of the lab.

The transformation from the Galactic frame to the equatorial frame is given by

\begin{equation}
\left( \begin{array}{c}
\hat{\mathbf{x}}_e \\
\hat{\mathbf{y}}_e \\
\hat{\mathbf{z}}_e \end{array} \right)=
\textbf{A}_\textrm{G}
\left( \begin{array}{c}
\hat{\mathbf{x}}_g \\
\hat{\mathbf{y}}_g \\
\hat{\mathbf{z}}_g \end{array} \right),
\end{equation}
where
\begin{equation}
\textbf{A}_\textrm{G}=
\left( \begin{array}{ccc}
a_x & b_x & c_x \\
a_y & b_y & c_y \\
a_z & b_z & c_z \end{array} \right),
\end{equation}
and $\hat{\mathbf{x}}_g$, $\hat{\mathbf{y}}_g$, and $\hat{\mathbf{z}}_g$ are the unit vectors of the Galactic reference frame. Recall the definition of the Galactic coordinate system: its origin is at the position of the Sun, its $x_g$-axis points towards the Galactic center, its $y_g$-axis points in the direction of the Galactic rotation, and its $z_g$-axis points to the north Galactic pole. Note that these coordinates are related
to Galactic longitude $l$ and latitude $b$ by ($x_g, y_g, z_g$)=($\cos b \cos l, \cos b \sin l, \sin b$). We have $\hat{\mathbf{q}}=q_{xg}~\hat{\bf x}_g + q_{yg}~\hat{\bf y}_g + q_{zg}~\hat{\bf z}_g$, where
\begin{align}
q_{xg} &= q_n \Big(-\big[a_x \cos (t^\circ_{\rm lab}) + a_y \sin (t^\circ_{\rm lab})\big] \sin(\lambda_{\rm lab}) +a_z\cos(\lambda_{\rm lab}) \Big) + q_w \Big(a_x \sin (t^\circ_{\rm lab}) -a_y \cos (t^\circ_{\rm lab}) \Big) \nonumber\\
&+ q_z \Big(\big[a_x \cos (t^\circ_{\rm lab}) +a_y \sin (t^\circ_{\rm lab})\big] \cos(\lambda_{\rm lab}) +a_z\sin(\lambda_{\rm lab}) \Big), \nonumber\\
q_{yg} &=q_n \Big(-\big[b_x \cos (t^\circ_{\rm lab}) + b_y \sin (t^\circ_{\rm lab})\big] \sin(\lambda_{\rm lab}) +b_z\cos(\lambda_{\rm lab}) \Big) + q_w \Big(b_x \sin (t^\circ_{\rm lab}) -b_y \cos (t^\circ_{\rm lab}) \Big) \nonumber\\
&+ q_z \Big(\big[b_x \cos (t^\circ_{\rm lab}) +b_y \sin (t^\circ_{\rm lab})\big] \cos(\lambda_{\rm lab}) +b_z\sin(\lambda_{\rm lab}) \Big),\nonumber\\
q_{zg} &=q_n \Big(-\big[c_x \cos (t^\circ_{\rm lab}) + c_y \sin (t^\circ_{\rm lab})\big] \sin(\lambda_{\rm lab}) +c_z\cos(\lambda_{\rm lab}) \Big) + q_w \Big(c_x \sin (t^\circ_{\rm lab}) -c_y \cos (t^\circ_{\rm lab}) \Big) \nonumber\\
&+ q_z \Big(\big[c_x \cos (t^\circ_{\rm lab}) +c_y \sin (t^\circ_{\rm lab})\big] \cos(\lambda_{\rm lab}) +c_z\sin(\lambda_{\rm lab}) \Big).
\label{qg}
\end{align}

As done in Ref.~\cite{Hipparcos}, the transformation matrix $\textbf{A}_\textrm{G}$ can be found using the definition of the Galactic pole and center in the \textit{International Celestial Reference System} (ICRS). The north Galactic pole can be defined by right ascension $\alpha_G=192^\circ .85948$ and declination $\delta_G=+27^\circ .12825$ in the ICRS. The origin of Galactic longitude is defined by the Galactic longitude of the ascending node of the Galactic plane on the equator of ICRS, which is $l_\Omega=32^\circ .93192$. The angles $\alpha_G$, $\delta_G$ and $l_\Omega$ should be regarded as exact quantities, and they can be used to compute the transformation matrix $\textbf{A}_\textrm{G}$~\cite{Hipparcos},
\begin{equation}
\textbf{A}_\textrm{G}=
\left( \begin{array}{ccc}
-0.0548755604 & +0.4941094279 & -0.8676661490\\
-0.8734370902 & -0.4448296300 & -0.1980763734\\
-0.4838350155 & +0.7469822445 & +0.4559837762 \end{array} \right).
\label{AG-Hipparcos}
\end{equation}

The direction of $\hat{\mathbf{q}}$ in the Galactic rest frame can be specified by the Galactic longitude ($l_q$) and latitude ($b_q$) of $\hat{\mathbf{q}}$,
\begin{equation}
q_{xg}=\cos{b_q}\cos{l_q},~~~~
q_{yg}=\cos{b_q}\sin{l_q},~~~~
q_{zg}=\sin{b_q},
\end{equation}
where $q_{xg}$, $q_{yg}$, and $q_{zg}$ are given in terms of $q_n$, $q_w$, and $q_z$ in Eq.~\ref{qg}.

The velocity of the lab  with respect to the center of the Galaxy can be divided into four components: the Galactic rotation velocity ${\bf V}_{\rm {Gal Rot}}$ at the position of the Sun (or Local Standard of Rest (LSR) velocity), Sun's peculiar velocity ${\bf V}_{\rm {Solar}}$ in the LSR, Earth's translational velocity ${\bf V}_{\rm {Earth Rev}}$ with respect to the Sun, and the  velocity of Earth's rotation around itself ${\bf V}_{\rm {Earth Rot}}$.
We also need to write each component of ${\bf V}_{{\rm lab}}$ in the Galactic frame and compute,
\begin{align}
\hat{\mathbf{q}} \cdot{\bf V}_{{\rm lab}}=\hat{\mathbf{q}} \cdot {\bf V}_{\rm {Gal Rot}}+ \hat{\mathbf{q}} \cdot {\bf V}_{\rm {Solar}}+ \hat{\mathbf{q}} \cdot {\bf V}_{\rm {Earth Rev}}+ \hat{\mathbf{q}} \cdot{\bf V}_{\rm {Earth Rot}}.
\label{qdotVlab}
\end{align}

\subsection{Galactic rotation}

The velocity of the Galactic rotation ${\bf V}_{\rm {Gal Rot}}$ is defined in the Galactic reference frame,
\begin{equation}
{\bf V}_{\rm {Gal Rot}}=V_{\rm {Gal Rot}} \hat{{\bf y}}_g.
\label{GalacticRot}
\end{equation}
Recall that $V_{\rm {Gal Rot}}$ is the Galactic rotation speed (i.e. the local circular speed), and $\hat{{\bf y}}_g$ is in the direction of the Galactic rotation. Thus we have
\begin{equation}
\hat{\mathbf{q}} \cdot{\bf V}_{\rm {Gal Rot}}=q_{yg} V_{\rm {Gal Rot}}
\label{qdotGalacticRot}
\end{equation}
where $q_{yg}$ is given in Eq.~\ref{qg}.

The standard value of $V_{\rm {Gal Rot}}$ for the IMB is 220 km/s~\cite{Kerr-1986}. As discussed in Ref.~\cite{Green:2010gw}, recent studies have found other values for  $V_{\rm {Gal Rot}}$. One analysis which was argued that it has used overly restrictive models found $V_{\rm {Gal Rot}}=(254 \pm 16)$ km/s~\cite{Reid-2009}. Another study found  $V_{\rm {Gal Rot}}=(236 \pm 11)$ km/s assuming a flat rotation curve~\cite{Bovy-2009}, while a different study found values ranging from $V_{\rm {Gal Rot}}=(200 \pm 20)$ km/s to $V_{\rm {Gal Rot}}=(279 \pm 33)$ km/s depending on the model used for the rotation curve~\cite{McMillan-2010}. We take $V_{\rm {Gal Rot}}=180$ km/s and 312 km/s as low and high estimates.

\subsection{Solar motion}

The Sun's peculiar velocity in the LSR is,
\begin{equation}
{\bf V}_{\rm {Solar}}=U \hat{{\bf x}}_g + V \hat{{\bf y}}_g +W \hat{{\bf z}}_g.
\label{Solar}
\end{equation}
Thus, we can compute $\hat{\mathbf{q}} \cdot{\bf V}_{\rm {Solar}}$ as
\begin{equation}
\hat{\bf q} \cdot{\bf V}_{\rm {Solar}}=q_{xg}U + q_{yg}V + q_{zg} W.
\label{qdotSolar}
\end{equation}

Ref.~\cite{Schoenrich-2010} has re-evaluated the velocity of the Sun with respect to the LSR and found that the classical determination of $U$, $V$, and $W$ is undermined by the metallicity gradient in the disc. In their study they find $(U,V,W)_\odot=(11.1^{+0.69}_{-0.75}, 12.24^{+0.47}_{-0.47}, 7.25^{+0.37}_{-0.36})$ km/s, with additional systematic uncertainties $\sim (1,2,0.5)$ km/s. The new values are extremely insensitive to the metallicity gradient within the disc.

\subsection{Earth's revolution}

The velocity of the Earth's revolution around the sun is given in terms of the Sun ecliptic longitude $\lambda(t)$ as~\cite{Lewin-1996}
\begin{align}
{\bf V}_{\rm {Earth Rev}}&=V_{\oplus}(\lambda) \Big[\cos\beta_x \sin (\lambda-\lambda_x) \hat{{\bf x}}_g + \cos\beta_y \sin (\lambda-\lambda_y) \hat{{\bf y}}_g + \cos\beta_z \sin (\lambda-\lambda_z) \hat{{\bf z}}_g\Big],
\label{EarthRev}
\end{align}
where $V_{\oplus}=29.8$ km/s is the orbital speed of the Earth, $V_{\oplus}(\lambda)=V_{\oplus}[1-e \sin(\lambda-\lambda_0)]$, $e=0.016722$, and $\lambda_0=13^\circ \pm 1^\circ$ are the ellipticity of the Earth's orbit and the ecliptic longitude of the orbit's minor axis, respectively, and $\beta_i=(-5^\circ.5303, 59^\circ.575, 29^\circ.812)$ and $\lambda_i=(266^\circ.141, -13^\circ.3485, 179^\circ.3212)$ are the ecliptic latitudes and longitudes of the ($\hat{{\bf x}}_g$,$\hat{{\bf y}}_g$,$\hat{{\bf z}}_g$) axes, respectively.

The Sun's ecliptic longitude $\lambda$ can be expressed as (Ref.~\cite{Green-2003} and p. 77  of Ref.~\cite{Lang}),
\begin{equation}
\lambda=L + (1^\circ .915 - 0^\circ.0048 T_0) \sin g+ 0^\circ .020 \sin 2g,
\end{equation}
where $L=281^\circ .0298 + 36000^\circ .77 T_0 + 0^\circ .04107 \textrm{UT}$ is the mean longitude of the Sun corrected for aberration, $g=357^\circ .9258 + 35999^\circ .05 T_0 + 0^\circ .04107 \textrm{UT}$ is the mean anomaly (polar angle of orbit). UT is the Universal Time in hours, and $T_0$ is the time in Julian centuries (36525 days) from 12:00 UT on 1 January 2010 to the previous midnight.

Thus, we can compute $\hat{\mathbf{q}} \cdot{\bf V}_{\rm {Earth Rev}}$ as
\begin{equation}
\hat{\mathbf{q}} \cdot{\bf V}_{\rm {Earth Rev}}=V_{\oplus}(\lambda) \Big[\cos\beta_x \sin (\lambda-\lambda_x) q_{xg} + \cos\beta_y \sin (\lambda-\lambda_y) q_{yg} + \cos\beta_z \sin (\lambda-\lambda_z) q_{zg}\Big].
\label{qdotEarthRev}
\end{equation}

\subsection{Earth's rotation}

Finally, we want to compute ${\bf V}_{\rm {Earth Rot}}$, the velocity of Earth's rotation around itself. We have
\begin{equation}
{\bf V}_{\rm {Earth Rot}}=-V_{\rm {RotEq}} \cos \lambda_{\rm lab} \hat{\cal W},
\end{equation}
where $V_{\rm {RotEq}}$ is the  Earth's rotation speed at the equator, and is defined as $V_{\rm {RotEq}}=2 \pi R_{\oplus}/({\rm {1~ sidereal~ day}})$. The Earth's equatorial radius is $R_{\oplus}=6378.137$ km, and one sidereal day is 23.9344696 hr$=86164$ s. therefore $V_{\rm {RotEq}}=0.465102$ km/s.

We use Eq. A8 of Ref.~\cite{ChanIV} to write $\hat{\cal W}$ in terms of the equatorial frame coordinates,
\begin{equation}
{\bf V}_{\rm {Earth Rot}}=-V_{\rm {RotEq}} \cos \lambda_{\rm lab} \Big(\sin (t^\circ_{\rm lab})\hat{\bf x}_e-\cos (t^\circ_{\rm lab})~\hat{\bf y}_e\Big),
\end{equation}
Using the the Galactic to equatorial transformation, we have
\begin{align}
{\bf V}_{\rm {Earth Rot}}&=-V_{\rm {RotEq}} \cos \lambda_{\rm lab}  \Big\{ \Big(a_x \sin (t^\circ_{\rm lab}) -a_y \cos (t^\circ_{\rm lab})\Big) \hat{{\bf x}}_g\nonumber\\
&+\Big(b_x\sin (t^\circ_{\rm lab}) -b_y \cos (t^\circ_{\rm lab}) \Big) \hat{{\bf y}}_g +\Big(c_x\sin (t^\circ_{\rm lab}) -c_y \cos (t^\circ_{\rm lab})\Big) \hat{{\bf z}}_g \Big\}.
\end{align}
Thus we can compute $\hat{\mathbf{q}} \cdot {\bf V}_{\rm {Earth Rot}}$ as
\begin{align}
\hat{\mathbf{q}} \cdot{\bf V}_{\rm {Earth Rot}}&=-V_{\rm {RotEq}} \cos \lambda_{\rm lab} \Big\{ \Big(a_x \sin (t^\circ_{\rm lab}) -a_y \cos (t^\circ_{\rm lab})\Big) q_{xg} \nonumber\\
&+\Big(b_x\sin (t^\circ_{\rm lab}) -b_y \cos (t^\circ_{\rm lab}) \Big) q_{yg} +\Big(c_x\sin (t^\circ_{\rm lab}) -c_y \cos (t^\circ_{\rm lab})\Big) q_{zg}\Big\}.
\label{qdotEarthRot}
\end{align}
Inserting $q_{xg}$, $q_{yg}$ and $q_{zg}$ from Eq.~\ref{qg}, we find
\begin{align}
\hat{\mathbf{q}} \cdot{\bf V}_{\rm {Earth Rot}}&=-V_{\rm {RotEq}} \cos \lambda_{\rm lab} \Big\{ \Big([a_y^2 +b_y^2 +c_y^2 -a_x^2 -b_x^2 -c_x^2] \cos (t^\circ_{\rm lab}) \sin (t^\circ_{\rm lab}) \sin \lambda_{\rm lab} \nonumber\\
&+[a_x a_y +b_x b_y +c_x c_y]\left(\cos^2 (t^\circ_{\rm lab}) -\sin^2 (t^\circ_{\rm lab})\right) \sin \lambda_{\rm lab}\nonumber\\
&+ [a_x a_z +b_x b_z + c_x c_z] \sin (t^\circ_{\rm lab}) \cos \lambda_{\rm lab} + [a_y a_z + b_y b_z + c_y c_z] \cos (t^\circ_{\rm lab}) \cos \lambda_{\rm lab} \Big) q_n \nonumber\\
&+\Big([a_x^2+b_x^2+c_x^2]\sin^2 (t^\circ_{\rm lab}) +[a_y^2+b_y^2+c_y^2] \cos^2 (t^\circ_{\rm lab})\nonumber\\
&-2[a_x a_y +b_x b_y +c_x c_y] \cos (t^\circ_{\rm lab}) \sin (t^\circ_{\rm lab}) \Big) q_w \nonumber\\
&+ \Big([a_x^2 +b_x^2 +c_x^2 -a_y^2 -b_y^2 -c_y^2] \cos (t^\circ_{\rm lab}) \sin (t^\circ_{\rm lab}) \cos \lambda_{\rm lab} \nonumber\\
&+[a_x a_y +b_x b_y +c_x c_y]\left(\sin^2 (t^\circ_{\rm lab}) -\cos^2 (t^\circ_{\rm lab})\right) \cos \lambda_{\rm lab}\nonumber\\
&+ [a_x a_z +b_x b_z + c_x c_z] \sin (t^\circ_{\rm lab}) \sin \lambda_{\rm lab} - [a_y a_z + b_y b_z + c_y c_z] \cos (t^\circ_{\rm lab}) \sin \lambda_{\rm lab} \Big) q_z  \Big\}.
\label{qdotEarthRot-expand}
\end{align}
From Eq.~\ref{AG-Hipparcos}, we have $a_i a_j + b_i b_j + c_i c_j=\delta_{ij}$ where $i,j=x,y,z$. Therefore, Eq.~\ref{qdotEarthRot-expand} will be simplified to
\begin{equation}
\hat{\mathbf{q}} \cdot{\bf V}_{\rm {Earth Rot}}=-q_wV_{\rm {RotEq}} \cos \lambda_{\rm lab}.
\label{qdotEarthRot-simplified}
\end{equation}

\subsection{Total Velocity}
Using Eqs.~\ref{qdotGalacticRot}, \ref{qdotSolar}, \ref{qdotEarthRev} and \ref{qdotEarthRot-simplified} we can compute $\hat{\mathbf{q}} \cdot{\bf V}_{{\rm lab}}$ as
\begin{align}
\hat{\mathbf{q}} \cdot{\bf V}_{{\rm lab}}&=q_{xg} \Big(U + V_{\oplus}(\lambda) \cos\beta_x \sin (\lambda-\lambda_x) \Big)+ q_{yg} \Big(V_{\rm {Gal Rot}} + V + V_{\oplus}(\lambda) \cos\beta_y \sin (\lambda-\lambda_y) \Big)\nonumber\\
&+ q_{zg} \Big(W + V_{\oplus}(\lambda) \cos\beta_z \sin (\lambda-\lambda_z) \Big) -q_wV_{\rm {RotEq}} \cos \lambda_{\rm lab},
\label{qdotVlab-2}
\end{align}
where $q_{xg}$, $q_{yg}$, and $q_{zg}$ can be written in terms of $q_n$, $q_w$, and $q_z$ using Eq.~\ref{qg}. We can also write $\hat{\mathbf{q}} \cdot{\bf V}_{{\rm lab}}$ in terms of only  $q_{xg}$, $q_{yg}$, and $q_{zg}$ using Eqs.~\ref{qdotGalacticRot}, \ref{qdotSolar}, \ref{qdotEarthRev} and \ref{qdotEarthRot},
\begin{align}
\hat{\mathbf{q}} \cdot{\bf V}_{{\rm lab}}&=q_{xg} \Big(U + V_{\oplus}(\lambda) \cos\beta_x \sin (\lambda-\lambda_x) -V_{\rm {RotEq}} \cos \lambda_{\rm lab} \left(a_x \sin (t^\circ_{\rm lab}) -a_y \cos (t^\circ_{\rm lab})\right)\Big) \nonumber\\
&+ q_{yg} \Big(V_{\rm {Gal Rot}} + V + V_{\oplus}(\lambda) \cos\beta_y \sin (\lambda-\lambda_y) -V_{\rm {RotEq}} \cos \lambda_{\rm lab} \left(b_x\sin (t^\circ_{\rm lab}) -b_y \cos (t^\circ_{\rm lab}) \right)\Big)\nonumber\\
&+ q_{zg} \Big(W + V_{\oplus}(\lambda) \cos\beta_z \sin (\lambda-\lambda_z) -V_{\rm {RotEq}} \cos \lambda_{\rm lab} \left(c_x\sin (t^\circ_{\rm lab}) -c_y \cos (t^\circ_{\rm lab})\right)\Big).
\label{qdotVlab-galactic}
\end{align}

\section{Logarithmic-ellipsoidal models}

In one logarithmic-ellipsoidal model provided by Evans, Carollo, and de Zeeuw~\cite{Evans:2000}, the Solar System is on the long (major) axis of the halo density ellipsoid, and
\begin{equation}
\sigma_{xg}^2=\frac{V_{\rm {Gal Rot}}^2}{(2+\Gamma)(\mathfrak{p}^{-2}+\mathfrak{q}^{-2}-1)},
\end{equation}
\begin{equation}
\sigma_{yg}^2=\frac{V_{\rm {Gal Rot}}^2 (2 \mathfrak{p}^{-2}-1)}{2(\mathfrak{p}^{-2}+\mathfrak{q}^{-2}-1)},
\end{equation}
\begin{equation}
\sigma_{zg}^2=\frac{V_{\rm {Gal Rot}}^2 (2 \mathfrak{q}^{-2}-1)}{2(\mathfrak{p}^{-2}+\mathfrak{q}^{-2}-1)}.
\end{equation}
In another model by the same authors, the Solar System is on the intermediate (minor) axis of the halo density ellipsoid, and
\begin{equation}
\sigma_{xg}^2=\frac{V_{\rm {Gal Rot}}^2 \mathfrak{p}^{-4}}{(2+\Gamma)(1+\mathfrak{q}^{-2}-\mathfrak{p}^{-2})},
\end{equation}
\begin{equation}
\sigma_{yg}^2=\frac{V_{\rm {Gal Rot}}^2 (2 - \mathfrak{p}^{-2})}{2(1+\mathfrak{q}^{-2}-\mathfrak{p}^{-2})},
\end{equation}
\begin{equation}
\sigma_{zg}^2=\frac{V_{\rm {Gal Rot}}^2 (2 \mathfrak{q}^{-2}-\mathfrak{p}^{-2})}{2(1+\mathfrak{q}^{-2}-\mathfrak{p}^{-2})}.
\end{equation}
$\mathfrak{p}$ and $\mathfrak{q}$ are constants used to describe the axis ratios of the density ellipsoid, and $\Gamma$ is a constant used to describe the velocity anisotropy.

We have studied 12 different combinations of parameters in the logarithmic-ellipsoidal model: 6 combinations for the Solar system on the major axis of the halo density ellipsoid and 6 on the minor axis. Eight of the cases are taken from Table II of Ref.~\cite{Alenazi-Gondolo:2008} and four from Table II of Ref.~\cite{Belli:2002}. The combinations of parameters for the Solar system on the major axis are
\begin{eqnarray}
\mathfrak{p}&=&0.9,~ \mathfrak{q}=0.8,~ \Gamma=-1.78,\nonumber\\
\mathfrak{p}&=&0.9,~ \mathfrak{q}=0.8,~ \Gamma=16,\nonumber\\
\mathfrak{p}&=&0.9,~ \mathfrak{q}=0.8,~ \Gamma=-1.0,\nonumber\\
\mathfrak{p}&=&0.9,~ \mathfrak{q}=0.8,~ \Gamma=-1.33,\nonumber\\
\mathfrak{p}&=&0.72,~ \mathfrak{q}=0.7,~ \Gamma=-1.39,\nonumber\\
\mathfrak{p}&=&0.72,~ \mathfrak{q}=0.7,~ \Gamma=-1.6,
\label{SunMajor}
\end{eqnarray}
and the combinations of parameters for the Solar system on the minor axis are
\begin{eqnarray}
\mathfrak{p}&=&0.9,~ \mathfrak{q}=0.8,~ \Gamma=-1.78,\nonumber\\
\mathfrak{p}&=&0.9,~ \mathfrak{q}=0.8,~ \Gamma=16,\nonumber\\
\mathfrak{p}&=&0.9,~ \mathfrak{q}=0.8,~ \Gamma=0.07,\nonumber\\
\mathfrak{p}&=&0.9,~ \mathfrak{q}=0.8,~ \Gamma=-0.62,\nonumber\\
\mathfrak{p}&=&0.72,~ \mathfrak{q}=0.7,~ \Gamma=4.02,\nonumber\\
\mathfrak{p}&=&0.72,~ \mathfrak{q}=0.7,~ \Gamma=2.01.
\label{SunMinor}
\end{eqnarray}


\begin{thebibliography}{99}

\bibitem{Drukier}
A. K. Drukier, K. Freese and D. N. Spergel, {\it Detecting cold dark-matter candidates}, Phys. Rev. D {\bf 33}, 3495 (1986);
K. Freese, J. Frieman and A. Gould, {\it Signal modulation in cold-dark-matter detection}, Phys. Rev. D {\bf 37}, 3388 (1988).

\bibitem{Ahlen:2009ev}
  S.~Ahlen {\it et al.},
  {\it The case for a directional dark matter detector and the status of current experimental efforts},
  Int.\ J.\ Mod.\ Phys.\  A {\bf 25}, 1-51 (2010)
  [arXiv:0911.0323 [astro-ph.CO]].


\bibitem{DRIFT}
G. J. Alner {\it et al.} [DRIFT Collaboration], {\it The DRIFT-I dark matter detector at Boulby: design, installation and operation}, Nucl. Instrum. Meth. A {\bf 535}, 644 (2004);
B. Morgan {\it et al.} [DRIFT Collaboration], {\it DRIFT: a directionally sensitive dark matter detector}, Nucl. Instrum. Meth. A {\bf 513}, 226 (2003).

\bibitem{DMTPC}
D. Dujmic {\it et al.} [DMTPC Collaboration], {\it Observation of the 'head-tail' effect in nuclear recoils of low-energy neutrons}, Nucl. Instrum. Meth. A {\bf 584}, 327 (2008);
A. Kaboth {\it et al.} [DMTPC Collaboration], {\it A Measurement of Photon Production in Electron Avalanches in CF4}, Nucl. Instrum. Meth. A {\bf 592}, 63 (2008).

\bibitem{NEWAGE}
K. Miuchi {\it et al.} [NEWAGE Collaboration], {\it Direction-sensitive dark matter search results in a surface laboratory}, Phys. Lett. B {\bf 654}, 58 (2007);
T. Tanimori {\it et al.}, {\it Detecting the WIMP-wind via spin-dependent interactions}, Phys. Lett. B {\bf 578}, 241 (2004).

\bibitem{MIMAC}
D. Santos {\it et al.}, {\it MIMAC: A Micro-TPC Matrix of Chambers for direct detection of Wimps}, J. Phys. Conf. Ser. {\bf 65}, 012012 (2007).


\bibitem{Spergel}
D. N. Spergel, {\it Motion of the Earth and the detection of weakly interacting massive particles},
Phys. Rev. D {\bf 37}, 1353 (1988).

\bibitem{Copi:1999pw}
  C.~J.~Copi, J.~Heo, L.~M.~Krauss,
  {\it Directional sensitivity, WIMP detection, and the galactic halo},
  Phys.\ Lett.\  B {\bf 461}, 43-48 (1999) [hep-ph/9904499];
  C.~J.~Copi, L.~M.~Krauss,
  {\it Angular signatures for galactic halo WIMP scattering in direct detectors: Prospects and challenges},
  Phys.\ Rev.\  D {\bf 63}, 043507 (2001)
  [astro-ph/0009467];
  B.~Morgan, A.~M.~Green,
  {\it Directional statistics for WIMP direct detection II: 2D read-out},
  Phys.\ Rev.\  D {\bf 72}, 123501 (2005)
  [astro-ph/0508134];
  C.~J.~Copi, L.~M.~Krauss, D.~Simmons-Duffin, S.~R.~Stroiney,
  {\it Assessing alternatives for directional detection of a wimp halo},
  Phys.\ Rev.\  D {\bf 75}, 023514 (2007)
  [astro-ph/0508649];
  A.~M.~Green, B.~Morgan,
  {\it The median recoil direction as a WIMP directional detection signal},
  Phys.\ Rev.\  D {\bf 81}, 061301 (2010)
  [arXiv:1002.2717 [astro-ph.CO]].

\bibitem{Morgan:2005}
B.~Morgan, A.~M.~Green, N.~J.~C.~Spooner,
  {\it Directional statistics for WIMP direct detection},
  Phys.\ Rev.\  D {\bf 71}, 103507 (2005)
  [astro-ph/0408047].

\bibitem{Green:2010gw}
  A.~M.~Green,
  {\it Dependence of direct detection signals on the WIMP velocity distribution},
  JCAP {\bf 1010}, 034 (2010) [arXiv:1009.0916 [astro-ph.CO]].


\bibitem{Billard:2009mf}
J. Billard, F. Mayet, J. F. Macias-Perez and D. Santos,
{\it Directional detection as a strategy to discover Galactic Dark Matter},
Phys. Lett. B {\bf 691}, 156 (2010) [arXiv:0911.4086 [astro-ph.CO]].


\bibitem{Gondolo:2002}
 P. Gondolo,
 {\it Recoil momentum spectrum in directional dark matter detectors},
 Phys.\ Rev.\ D {\bf 66}, 103513 (2002).


\bibitem{ChanIV}
N. Bozorgnia, G. Gelmini and P. Gondolo,
{\it Daily modulation due to channeling in direct dark matter crystalline detectors},
  Phys. Rev. D {\bf 84}, 023516 (2011) [arXiv:1101.2876 [astro-ph.CO]].

\bibitem{Alenazi-Gondolo:2008} M. S. Alenazi and P. Gondolo,
{\it Directional recoil rates for WIMP direct detection},
Phys.\ Rev.\ D {\bf 77}, 043532 (2008).


 \bibitem{Kuhlen} M. Kuhlen {\it et al.},
  {\it Dark Matter Direct Detection with Non-Maxwellian Velocity Structure},
JCAP {\bf 1002}, 030 (2010) [arXiv:0912.2358v1 [astro-ph.GA]].

\bibitem{RAVE}
 M. C. Smith {\it et al.},
 {\it The RAVE Survey: Constraining the Local Galactic Escape Speed},
 Mon. Not. Roy. Astron. Soc. {\bf 379}, 755 (2007) [astro-ph/0611671].


\bibitem{HEALPix:2005} K. M. G\'{o}rski {\it et al.},
{\it HEALPix: A Framework for High-Resolution Discretization and Fast Analysis of Data Distributed on the Sphere},
 ApJ {\bf 622}, 759 (2005).


\bibitem{Helm:1956} R. Helm,
{\it Inelastic and elastic scattering of 187-Mev electrons from selected even-even nuclei},
Phys.\ Rev.\ {\bf 104}, 1466 (1956).

\bibitem{Weber:2010} M. Weber and W. de Boer,
{\it Determination of the Local Dark Matter Density in our Galaxy},
Astron. Astrophys. {\bf 509}, 25 (2010) [arXiv:0910.4272].



\bibitem{Salucci:2010}
  P.~Salucci, F.~Nesti, G.~Gentile, C.~F.~Martins,
  {\it The dark matter density at the Sun's location},
  Astron.\ Astrophys.\  {\bf 523}, A83 (2010).
  [arXiv:1003.3101 [astro-ph.GA]].

\bibitem{Pato:2010}
  M.~Pato {\it et al.}
  {\it Systematic uncertainties in the determination of the local dark matter density},
  Phys.\ Rev.\  D {\bf 82}, 023531 (2010).
  [arXiv:1006.1322 [astro-ph.HE]].

\bibitem{Catena:2010}
  R.~Catena, P.~Ullio,
  {\it A novel determination of the local dark matter density},
  JCAP {\bf 1008}, 004 (2010) [arXiv:0907.0018 [astro-ph.CO]].

\bibitem{Gelmini-Gondolo:2001} G. Gelmini and P. Gondolo,
{\it WIMP Annual Modulation with Opposite Phase in Late-Infall Halo Models},
Phys. Rev. D {\bf 64}, 023504 (2001) [arXiv:hep-ph/0012315]


\bibitem{Green:2008}
A. M. Green, {\it Determining the WIMP mass from a single direct detection experiment, a more detailed study},
JCAP {\bf 0807}, 005 (2008) [arXiv:0805.1704v2 [hep-ph]].

\bibitem{Shan:2009}
C.L. Shan,
{\it Determining the mass of dark matter particles with direct detection experiments},
New J. Phys. {\bf 11}, 105013 (2009).


\bibitem{Schnee:2011}
R. W. Schnee,
{\it Introduction to Dark Matter Experiments},
In Physics of the Large and Small: Proceedings of the 2009 Theoretical Advanced Study Institute in Elementary Particle Physics, pp. 629-681 (World Scientific, Singapore), Ed. Csaba Csaki and Scott Dodelson (2010)
[arXiv:1101.5205v1 [astro-ph.CO]].

\bibitem{Billard:2011}
J. Billard, F. Mayet and D. Santos,
{\it A Markov chain Monte Carlo analysis to constrain dark matter properties with directional detection},
Phys. Rev. D {\bf 83}, 075002 (2011) [arXiv:1012.3960v2 [astro-ph.CO]].

\bibitem{Lee:2012}
S. K. Lee and A. H. G. Peter,
{\it Probing the local velocity distribution of WIMP dark matter with directional detectors},
JCAP {\bf 1204}, 029 (2012) [arXiv:1202.5035v1 [astro-ph.CO]].


\bibitem{Moore:2001}
B. Moore {\it et al.},
{\it Dark matter in Draco and the Local Group: Implications for direct detection experiments},
Phys. Rev. D {\bf 64}, 063508 (2001).

\bibitem{Helmi:2002}
A. Helmi, S. D. M. White, and V. Springel,
{\it The phase-space structure of a dark-matter halo: Implications for dark-matter direct detection experiments},
Phys. Rev. D {\bf 66}, 063502 (2002).

\bibitem{Green:2002}
A. M. Green,
{\it Effect of halo modeling on weakly interacting massive particle exclusion limits},
Phys. Rev. D {\bf 66}, 083003 (2002).

\bibitem{Evans:2000}
N. W. Evans, C. M. Carollo, and P. T. de Zeeuw,
{\it Triaxial haloes and particle dark matter detection},
Mon. Not. R. Astron. Soc. {\bf 318}, 1131 (2000).



  \bibitem{Hipparcos}
The Hipparcos and Tycho Catalogues, Volume 1, Section 1.5, \textit{Transformation of Astrometric Data
and Associated Error Propagation}, ESA Publications Division, c/o ESTEC, Noordwijk, The Netherlands, 1997,
http://www.rssd.esa.int/SA-general/Projects/Hipparcos/ CATALOGUE\_VOL1/sect1\_05.pdf.


  \bibitem{Kerr-1986}
  F. J. Kerr and D. Lynden-Bell,
  {\it Review of galactic constants},
  Mon. Not. Roy. Astron. Soc. {\bf 221}, 1023 (1986).


\bibitem{Reid-2009}
M. J. Reid et al.,
{\it Trigonometric Parallaxes of Massive Star Forming Regions: VI. Galactic Structure, Fundamental Parameters and Non-Circular Motions},
Astrophys. J. {\bf 700}, 137 (2009) [arXiv:0902.3913].

 \bibitem{Bovy-2009}
J. Bovy, D. W. Hogg and H. Rix,
{\it Galactic masers and the Milky Way circular velocity},
Astrophys. J. {\bf 704}, 1704 (2009) [arXiv:0907.5423].

\bibitem{McMillan-2010}
P. J. McMillan and J. J. Binney,
{\it The uncertainty in Galactic parameters},
Mon. Not. Roy. Astron. Soc. {\bf 402}, 934 (2010) [arXiv:0907.4685].

\bibitem{Schoenrich-2010}
R. Schoenrich, J. Binney and W. Dehnen,
{\it Local Kinematics and the Local Standard of Rest},
Mon. Not. Roy. Astron. Soc. {\bf 403} 1829 (2010) [arXiv:0912.3693].

\bibitem{Lewin-1996}
J.D. Lewin and P.F. Smith,
{\it Review of mathematics, numerical factors, and corrections for dark matter experiments based on elastic nuclear recoil},
Astropart. Phys. {\bf 6}, 87 (1996).

\bibitem{Green-2003} A. M. Green,
{\it Effect of realistic astrophysical inputs on the phase and shape of the WIMP annual modulation signal},
Phys. Rev. D {\bf 68}, 023004 (2003) [arXiv:astro-ph/0304446].

\bibitem{Lang} K. R. Lang, {\it Astrophysical Formulae}, Springer-Verlag, New York (1999).


\bibitem{Belli:2002}
P. Belli, R. Cerulli, N. Fornengo, and S. Scopel,
{\it Effect of the galactic halo modeling on the DAMA/NaI annual modulation result: an extended analysis of the data for WIMPs with a purely spin-independent coupling},
Phys. Rev. D {\bf 66}, 043503 (2002) [arXiv:hep-ph/0203242].



\end{thebibliography}
\end{document}